\documentclass[referee,pdflatex,sn-standardnature]{sn-jnl}

\jyear{2021}%
\usepackage{amsmath} 

\usepackage{natbib}
\usepackage{url}
\usepackage{mwe}
\usepackage[markcase=noupper
]{scrlayer-scrpage}
\ohead{}
\theoremstyle{thmstyleone}%
\theoremstyle{thmstyletwo}%
\usepackage[modulo]{lineno}

\theoremstyle{thmstylethree}%

\unnumbered
\begin{document}

\title[Article Title]{Anisotropic positive linear and sub-linear magnetoresistivity in the cubic type-II Dirac metal Pd$_{3}$In$_{7}$}


\author[1,2]{\fnm{Aikaterini} \sur{Flessa Savvidou}}

\author[3]{\fnm{Andrzej} \sur{Ptok}}
\author[4]{\fnm{G.} \sur{Sharma}}
\author[1]{\fnm{Brian} \sur{Casas}}
\author[5]{\fnm{Judith K.} \sur{Clark}}
\author[5]{\fnm{Victoria M.} \sur{Li}}
\author[5]{\fnm{Michael} \sur{Shatruk}}
\author[6]{\fnm{Sumanta} \sur{Tewari}}
\author*[1,2]{\fnm{Luis} \sur{Balicas}}\email{balicas@magnet.fsu.edu}

\affil[1]{\orgdiv{National High Magnetic Field Laboratory}, \orgname{Florida State University}, \orgaddress{ \city{Tallahassee}, \postcode{32310}, \state{Florida}, \country{USA}}}

\affil[2]{\orgdiv{Department of Physics}, \orgname{Florida State University}, \orgaddress{ \city{Tallahassee}, \postcode{32306}, \state{Florida}, \country{USA}}}

\affil[3]{\orgdiv{Institute of Nuclear Physics}, \orgname{Polish Academy of Sciences}, \orgaddress{\street{W. E. Radzikowskiego 152}, \city{Krak\'{o}w}, \postcode{PL-31342}, \country{Poland}}}

\affil[4]{\orgdiv{School of Physical Sciences}, \orgname{Indian Institute of Technology Mandi}, \orgaddress{ \city{Mandi}, \postcode{175005}, \state{H.P.}, \country{India}}}

\affil[5]{\orgdiv{Department of Chemistry and Biochemistry}, \orgname{Florida State University}, \orgaddress{ \city{Tallahassee}, \postcode{32306}, \state{Florida}, \country{USA}}}

\affil[6]{\orgdiv{Department of Physics and Astronomy}, \orgname{Clemson University}, \orgaddress{ \city{Clemson}, \postcode{29634}, \state{South Carolina}, \country{USA}}}


\abstract{We report a transport study on Pd$_3$In$_7$ which displays multiple Dirac type-II nodes
in its electronic dispersion. Pd$_3$In$_7$ is characterized by low residual resistivities and high mobilities, which are consistent with  Dirac-like quasiparticles. For an applied magnetic field $(\mu_{\text{0}} H)$ having a non-zero component along the electrical current, we find a large, positive, and linear in  $\mu_{\text{0}} H$ longitudinal magnetoresistivity (LMR). The sign of the LMR and its linear dependence deviate from the behavior reported for the chiral-anomaly-driven LMR in Weyl semimetals. Interestingly, such anomalous  LMR is consistent with predictions for the role of the anomaly in type-II Weyl semimetals.  In contrast, the transverse or conventional magnetoresistivity (CMR for electric fields $\textbf{E} \bot \mu_{\text{0}} \textbf{H}$) is  large and positive, increasing by $10^3-10^4$ \% as a function of $\mu_{\text{0}}H$ while following an anomalous, angle-dependent power law $\rho_{\text{xx}}\propto (\mu_{\text{0}}H)^n$
with $n(\theta) \leq 1$. The order of magnitude of the CMR, and its anomalous power-law, is explained in terms of uncompensated electron and hole-like Fermi surfaces characterized by anisotropic carrier scattering likely due to the lack of Lorentz invariance.}

\maketitle

\section{\textbf{INTRODUCTION}}

Triggered by the discovery of topological insulators \cite{RevModPhys.83.1057}, the condensed matter community has, in recent years, focused intensively on the topological nature of the electronic band structure of materials. The subsequent discovery of Dirac \cite{Na3BiPRB,Na3Bi} and Weyl semimetals \cite{TaAs,TaAs_Science} as well as higher order topological systems \cite{Higher_order_TIs} continues to strengthen this interest. In particular, the study of Dirac materials  is at the forefront of research, ranging from $d$-wave superconductors \cite{RevModPhys.78.373}, graphene \cite{RevModPhys.81.109}, bulk semi-metallic systems such as Cd$_3$As$_2$ \cite{LiangT}, to Lorentz invariance violating systems that are characterized by tilted Dirac/Weyl cones, or the so-called type-II Dirac/Weyl compounds. \cite{Type-II_Dirac,Type-II_Dirac_PRL,PtTe2}.

This intense interest in topological compounds relies on the prediction and observation of unreported phenomena such as a fermion chirality dependent circular galvanometric effect \cite{Ma_CPGE,Wu_CPGE,RhSi_CPGE}, a Hall-effect in absence of time-reversal symmetry \cite{WTe2_Ma,Ma1}, or the prediction for an anomalous magnetoresistivity resulting from the chiral anomaly among Weyl nodes \cite{Adler,Bell_Jackiw,Pallab,Fang,Jia,Ong,Burkov}.
In conventional or three-dimensional type-I Dirac/Weyl systems, the chiral or axial anomaly is predicted \cite{Adler,Bell_Jackiw,Pallab,Fang,Jia,Ong,Burkov} to provide an additional positive contribution to the longitudinal magneto-conductivity (LMC) of semimetallic systems when a component of the external magnetic field $\mu_0\bf{H}$ is aligned along the electric field $\bf{E}$ driving the current density $\bf{j}$. Essentially, $\mu_0\bf{H}$ favors one fermion chirality to the detriment of the other inducing a net charge transfer, or an axial current between nodes.

In type-II Dirac/Weyl systems, the existence of tilted Dirac/Weyl cones also leads to an anomalous contribution to the LMC \cite{Bernevig}. It was originally predicted \cite{Bernevig} to occur when $\mu_0\bf{H}$ is oriented within a cone in $k-$space satisfying the condition
$\lvert T(\bf k)\rvert >$ $ \lvert U(\bf k)\rvert$, where $\lvert T(\bf k)\rvert$ and $\lvert U(\bf k) \rvert$ represent the kinetic and potential energy components of the linear  energy dispersion. Subsequent work found that this anomaly introduces a linear in field contribution to the LMC for fields applied along the tilt direction, but a quadratic one under perpendicularly applied fields~\cite{sharma2017chiral,zyuzin2017magnetotransport}. Recent studies paint a more nuanced and complex scenario, with the anomaly leading even to a positive contribution to the LMR, in contrast to the proposed positive contributions to the LMC \cite{Kondo,Annals,ahmad2021longitudinal}. Furthermore, according to Refs. \cite{Kondo,Annals,ahmad2021longitudinal} the observation of positive versus negative LMR would depend on the tilt direction of the Dirac/Weyl cones, the level of tilting, their relative inclination, relative separation, the orientation of $\mu_0\bf{H}$ relative to the vector connecting them, and the strength of intervalley scattering. Throughout this text, conventional magnetoresistivity refers to a configuration where $ \bf{j}$  $\bot$  $\mu_0 \bf{H}$

Type-II Weyl semimetals provide another interesting scenario due to the presence of both electron and hole pockets coexisting with the Weyl points on the Fermi surface. As observed in the type-II Weyl semimetal WTe$_2$~\cite{WTe2}, a near exact compensation of electron and hole pockets may result in a very large and non-saturating magnetoresistivity that varies quadratically with the applied field. Here, we shall show that even uncompensated electron and hole pockets can yield significant CMR and exhibit an anomalous power law dependence on the magnetic field: $\rho_{\text{xx}}\propto (\mu_0H)^{n}$, with $n \leq 1$.  We thereby investigate both the unconventional LMR  and the  anomalous CMR that result from the axial anomaly acting on a type-II Dirac/Weyl system that lacks carrier compensation and is characterized by anisotropic carrier scattering on its Fermi surfaces.

To understand the influence of electronic topology on the magnetoresistivity, a number of Dirac/Weyl systems have been recently studied, including proposed type-II Weyl semimetals such as WTe$_2$ \cite{WTe2} or NbP \cite{NbP}, three-dimensional type-I Dirac semimetals such as Cd$_3$As$_2$ \cite{LiangT} or Na$_3$Bi \cite{Na3Bi}, type-I Weyl semimetallic systems such as TaAs \cite{TaAs,TaAs_Science} and many others. These materials exhibit high carrier mobilities \cite{LiangT}, in some cases very large, and linear CMR \cite{LiangT, WTe2, CoS2} and evidence for the chiral anomaly (negative LMR) \cite{Na3Bi,TaAs,TaAs_Science}. Experimentally, there have been few reports on the transport properties of pure type-II Dirac systems, with the CMR of systems like NiTe$_2$ or SrAgB displaying  linear \cite{Nite2_Xu, SrAgB}, and in occasions a sublinear \cite{NiTe2} dependence on field. And systems displaying substantial residual resistivities, i.e., tens of $\mu \Omega$ cm like Ir$_2$In$_8$S, claimed to display power laws in field with  exponents ranging between 1 and 2 \cite{Ir2In8S}. It is therefore important to expose the transport properties of very clean and isotropic type-II Dirac systems that do not display any type of ordering.

Here, we focus on compounds belonging to the Ir$_3$Ge$_7$ family of cubic structures that are known to display a remarkable compositional variability. Chemical substitutions can vary the electron count widely leading to semiconductors like Mo$_3$Sb$_5$Te$_2$ \cite{Mo3Sb5Te2}, metals like Pd$_3$In$_7$ and even superconductors such as Mo$_3$Sb$_7$ \cite{Mo3Sb7}.
Metallic compounds like Pd$_3$In$_7$ or Pt$_3$In$_7$ have already been previously predicted to be topologically non-trivial \cite{Bernevig1}. Furthermore, the electronic band structure of the newly discovered Rh$_3$In$_{3.4}$Ge$_{3.6}$ compound, which belongs to the same structural family, displays multiple band crossings close to the Fermi energy $\varepsilon_{\text{F}}$, leading to type-I, type-II and even type-III Dirac nodes \cite{Katerina}. Our calculations reveal the existence of multiple type-II Dirac nodes near the Fermi level of Pd$_3$In$_7$. In this study, we find Pd$_3$In$_7$ to display a very low residual resistivity, high carrier mobilities due to low carrier effective masses, and pronounced CMR, which are the transport hallmarks of Dirac-like quasiparticles. Remarkably, this compound tends to display linear CMR but with a superimposed conventional quadratic in-field term \cite{Pippard} at rather low fields. This dependence crosses over to an angle dependent power law $\rho_{\text{xx}}\propto (\mu_0H)^{n}$  at higher fields, with $n < 1 $, except for a very specific magnetic field orientation where $n = 1 $ over the entire field range. In a LMR configuration, that is for $\mathbf{j} \| \mu_0\mathbf{H}$, we do not observe the conventional saturating magnetoresistivity \cite{Pippard1}. Instead, a positive LMR is observed under fields as high as $\mu_0H$ = 35 T.

Given that Pd$_3$In$_7$ is a clean system that remains well below the quantum limit under available magnetic fields, and displays no magnetic or electronic order, we  conclude that its anomalous LMR  results from the axial anomaly among type-II Dirac/Weyl nodes.
In contrast, the sublinear dependence of the CMR on the magnetic field is explained as a multi-band effect due to the presence of uncompensated electron and hole pockets of distinct carrier mobilities. The angular dependence of the power-law exponent as the field is rotated, is attributed to a pronounced angle-dependent Zeeman effect resulting from the large and anisotropic value of the Land\'{e} $g$-factor in Pd$_3$In$_7$. This is likely to lead to an anisotropic modification of the geometry of the Fermi surface(s) changing the scattering rates or mobilities of electrons and holes on them, as the field is rotated relative to the crystallographic axes. This effect might be pronounced on the smallest Fermi surfaces. We argue that this might explain the experimentally observed angular dependence of the exponent $n(\theta)$ in an uncompensated metal like Pd$_3$In$_7$.

\section{\label{sec:level2}RESULTS}
\subsection{Anomalous Magnetoresistivity in \texorpdfstring{Pd$_3$In$_7$}{Lg}}

Figures 1a and 1b provide a schematic depicting the planes of rotation of the magnetic field relative to the crystallographic axes, as well as the definition of the angles $\phi$ and $\theta$, used for measuring the longitudinal and conventional magnetoresistivities, respectively. The planes of rotation were defined by the natural morphology of the measured single-crystals.
The resistivity $\rho_{\text{xx}}$ as a function of the temperature $T$ for a Pd$_3$In$_7$ single crystal is displayed in Fig. 1c, revealing metallic behavior with a linear dependence on $T$ (red line is a linear fit) for $T > 50$ K. For $T \leqq 50$ K, it follows a quadratic dependence on $T$ (cyan  line is a fit to a $T^2$ term), indicating Fermi-liquid behavior. The high value of the residual resistivity ratio, $RRR = (\rho_{\text{xx}}(300K)- \rho_{\text{xx}}(2K))/ \rho_{\text{xx}}(2K) \simeq 137$, as well as the low residual resistivity $\rho_0 \simeq 110$ n$\Omega$ cm (inset in Fig. 1c), point to the very high crystalline quality of the sample, or its low level of disorder. To evaluate the mean transport mobility $\mu_{\text{tr}}$ of our crystals, we calculated the Hall conductivity $\sigma_{\text{xy}}$ based on the magnetoresistivity $\rho_{\text{xx}}$ and the Hall resistivity $\rho_{\text{xy}}$:
\begin{equation}
\sigma_{\text{xy}}= - \frac{\rho_{\text{xy}}}{\rho_{\text{xx}}^2+\rho_{\text{xy}}^2}
\end{equation}
    $\sigma_{\text{xy}}$ as a function of $\mu_0H$ is plotted in Fig. 1d. From $\sigma_{\text{xy}}$, we can extract the mean transport mobility of the charge carriers: $\mu_{\text{tr}}=(\mu_0 H_0)^{-1}=1.2 \times 10^4$\text{ cm}$^2$V$^{-1}$s$^{-1}$. This value is comparable to those extracted from type-II Dirac semimetallic systems like MoTe$_2$ \cite{Qiong} or WP$_2$ \cite{Rico}, which are characterized by extremely high magnetoresistivities, but considerably smaller  than those values reported for Dirac nodal line systems like ZrSiSe \cite{Michael}, and over two orders of magnitude smaller than the mobilities reported for the Dirac semimetal  Cd$_3$As$_2$ \cite{LiangT}. In contrast, Pd$_3$In$_7$ is a metal characterized by large Fermi surfaces (or large carrier densities) as discussed below, or is not a nearly compensated semimetal as is the case for those systems. The mean free path $l_0 = v_{\text{F}} \mu_{\text{tr}} m^{\star}/e $ ($v_{\text{F}}$ is the Fermi velocity and $e$ the electron charge) can be estimated from the effective mass $m^{\star}$ extracted from the de Haas-van-Alphen (dHvA) oscillations (discussed below) and the average transport mobility $\mu_{\text{tr}}$, yielding Fermi surface cross-sectional area dependent values $l_0 \approx 0.88 - 4.75$ $\mu$m at $T=0.35$ K. The low value of the residual resistivity  $\rho_0$ and the large values of $l_0$ clearly indicate that impurities have a negligible role on the transport properties of Pd$_3$In$_7$. This is an important point for the subsequent discussions.

The magnetic field dependence of the magnetoresistivity MR depends on the angle of rotation relative to the main crystallographic axes. The MR, in $\%$, is defined as:
\begin{equation}
\text{MR}  = \frac{\rho_{\text{xx}}(\mu_0H)-\rho_{\text{xx}}(0)}{\rho_{\text{xx}}(0)}\cdot 100
\end{equation}

As depicted in Fig. 1a to measure LMR, the magnetic field $\mu_0H $ was rotated from an initial position perpendicular to the current $I$ or $\phi=0^{\circ}$, with $\mu_0\textbf{H} \parallel (-1,2,-1)$, and towards  $\mu_0\textbf{H} \parallel \textbf{I} \parallel (-1,0,1)$ or $\phi=90^{\circ}$. For the conventional magnetoresistivity or CMR, $\mu_0H $ remains perpendicular to the current $I$ over the entire angular range, with  $\theta=0^{\circ}$ corresponding to $\mu_0\textbf{H} \parallel (-1,2,-1)$ and $\theta=90^{\circ}$ to $\mu_0\textbf{H} \parallel (1,1,1)$ (see, Fig. 1b).  The magnetic field was rotated within these specific crystallographic planes due to the morphology of the as-grown crystals. Electrical contacts were placed on the largest as-grown crystalline surface. Previously, the crystal was  polished on the opposite surface to decrease its thickness. In Pd$_3$In$_7$ crystals, planes perpendicular to the main crystallographic axes, i.e., (1,0,0), (0,1,0) or (0,0,1), yield rather small surfaces which precluded transport measurements with magnetic fields applied along them. The lowest temperatures were chosen to minimize the role of phonons.

As seen in the inset of Fig. 1e, most field orientations lead to a similar MR behavior, quadratic in field dependence at low fields and sublinear at higher fields (Fig. 1e, black line). A sublinear MR in a very clean system, that displays several Dirac crossings in its band structure (as we show below), has not been thoroughly exposed and characterized.
In the inset of Fig. 1e, or for $\theta=35^{\circ}$,  the quadratic in field response is observed for $\mu_0H <$ 4 T where the MR is fitted (red line) to a combination of linear and quadratic components: MR $ = d_1 + d_2  (\mu_0 H)+ d_3 (\mu_0 H)^2$. At the same angle and for $\mu_0H > 4$ T, the MR becomes sub-linear in $\mu_0H$, but as seen in Fig. 1e for the black trace, it does not saturate  even under fields as large as $\mu_0H$ = 41.5 T. The red dashed line is a guide to eye indicating the deviation of the raw data with respect to linear behavior.

Over a broad range of fields the MR can be fitted to a power law (red line in Supplementary Fig. 4), i.e., MR = $a(\mu_0 H)^n$ with $n < 1$. Remarkably, and as seen in Fig. 1e, blue line, when the angle of rotation is $\theta=0^{\circ}$ the MR displays a linear dependence on $\mu_0H$ from $\mu_0H = 0$ T all the way up to $\mu_0H = 41.5$ T. The magenta dashed line is a linear fit illustrating this point. At the highest fields, the observed deviations with respect to linearity are attributable to Shubnikov-de Haas oscillations.  This linear dependence is observed only at this precise orientation, whereas for all other angles the MR displays a combination of linear and quadratic terms for $\mu_0H < 4$ T,  followed by sublinear dependence at higher fields. Notice that the linear MR of Pd$_3$In$_7$ is far more pronounced than that of Ag$_{2+\delta}$Se, a compound previously proposed as a possible megaGauss sensor \cite{AgSe}, and extends to higher fields.
This linearity extends to higher temperatures and follows Kohler's scaling \cite{Kohler}, as seen in Supplementary Fig. 5. The LMR displays a positive, and nearly linear in field behavior over the entire $\phi$ range. In Supplementary Fig. 6, the LMR is displayed for $\mu_0\textbf{H} \parallel \textbf{I}$, or $\phi = 90^{\circ}$.

This linear (and sublinear) in magnetic field dependence for the CMR  contrasts markedly with previous reports for type-I Weyl semimetals, i.e, a $(\mu_0H)^2$ dependence followed by a linear one at high fields \cite{Minhyea}.  For compensated semimetals, the dependence on the magnetic field is quadratic over the entire field range \cite{WTe2,Minhyea}. As we discuss below, the linear as well as sublinear dependence on the magnetic field is explained as a multi-band effect due to the presence of uncompensated electron and hole pockets in the Type-II Dirac metal Pd$_3$In$_7$.

For an axis of rotation that maintains the magnetic field perpendicular to the electrical current, i.e., for a configuration that maintains the  Lorentz force at its maximum value throughout the entire angular range, we observe an anisotropic four-fold, butterfly-like, angular MR (Figs. 2a and 2b). Figure 2b shows the CMR for various magnetic fields, showing an increase of $\sim 1000$ $\%$ under $\mu_0H$ = 9 T. Figure 2a corresponds to a polar plot of the CMR in units of $\%$. From the polar plot one observes two- and fourfold symmetries, as well as dips in the CMR at $\theta = 0^{\circ}$, $30^{\circ}$ and $90^{\circ}$. It is important to emphasize that this angular structure cannot be attributed to the superimposed Shubnikov-de Haas oscillations. Supplementary Fig 1 displays the CMR as a function of $\mu_0H$ for several values of $\theta$. At the highest fields, its background, upon which the oscillations are superimposed, displays two clear minima at $\theta \sim 30^{\circ}$ and $\theta \sim 55^{\circ}$. This confirms the intrinsic origin of the angular structure in the CMR that originates  the butterfly like magnetoresistivity seen at lower fields, albeit over the entire angular range the current is always maintained perpendicular to the field.

For a rotation starting from $\mathbf{I}\perp \mu_0 \mathbf{H}$ or $\phi = 0^{\circ}$ tilting towards $\phi = 90^{\circ}$ or $\mathbf{I}\parallel \mu_0 \mathbf{H}$, thus modulating the strength of the Lorentz force, the MR is shown in Fig. 2c. As expected, for $\phi = 0^{\circ}$ the MR reaches its maximum value of $7500\%$ under $\mu_0H = 35$ T,  decreasing as the angle increases or as the Lorentz force decreases. Nevertheless, in the region between $\phi = 30^{\circ}$ and $\phi = 55^{\circ}$, the MR displays a second maximum whose origin remains to be understood.
This so-called butterfly magnetoresistivity, was previously reported for the ZrSiS family of compounds and attributed to charge carriers initially exploring  topologically trivial orbits on the Fermi surface that become non-trivial as the magnetic field rotates with respect to the main crystallographic axes \cite{Mazhar}.
Subsequent work concluded the existence of an anisotropic scattering rate on its Fermi surface that is probed by the electronic orbits upon rotation of the magnetic field \cite{Michael}.

To further expose the sublinear magnetoresistivity in  Pd$_3$In$_7$, we plot in Figs. 3a and 3b  the power exponent $n$ as well as the coefficient $\alpha$ respectively, extracted by fitting the raw MR as a function of $\mu_0H$ (Fig. 1e), measured at several angles, to the function MR $ = \alpha (\mu_0H)^n$. For $\theta =0^{\circ}$, we obtain $n = 1$  but it progressively decreases as the $\theta$ increases, displaying a minimum $n \simeq 0.7$ around $\theta = 55^{\circ}$ - $60^{\circ}$. In contrast, the coefficient $\alpha$ displays a maximum around this angular range, increasing by nearly $250\%$ relative to its value at $0^{\circ}$. The power law exponent was also calculated through the derivative of $\rho_{\text{xx}} =  \rho_0 + \alpha (\mu_0 H)^n$, i.e., $n = \partial\ln(\rho_{\text{xx}} - \rho_0)/\partial\ln(\mu_0H)$, in order to expose its field and angular dependence. In Figs. 3c and 3d, we plot $n$  for two distinct crystallographic planes of rotation  of $\mu_0H$ relative to $I$. In Fig. 3c, $\mathbf{I}$ is kept $\perp \mu_0 \mathbf{H}$ while $\mu_0\textbf{H}$ rotates, or the angle $\theta$ is varied, as seen in Fig. 1b. The plot  reveals $n=2$ for $\mu_0H < 4$ T, which becomes $n \sim 1$ for $4 \leq \mu_0H \leq 10$ T. Above 10 T, $n$ takes values $0.5 \leqq n \leqq 0.9$, for all angles excluding $\theta=0^{\circ}$. In contrast, when the current is rotated from $\mathbf{I}\perp \mu_0 \mathbf{H}$ to $ \mathbf{I}\parallel \mu_0 \mathbf{H}$, $n \approx 1$ is extraced for the entire angular range, see Fig. 3d. In this panel, deviations with respect to $n=1$ result from noise in the derivative due, for example, to the superimposed Shubnikov-de Haas oscillations.  Both sets of measurements were performed on the same crystal, with the electrical contacts attached to the same crystallographic plane.  The raw magnetoresisistivity data $\rho_{xx}(\mu_0H)$ for multiple values of both angles $\phi$ and $\theta$, are presented in Supplementary Figs. 1 and 2 in the SI file. Small differences between distinct $\theta=0^{\circ} $and $\phi=0^{\circ}$ traces shown in the main text and in the SI file, are attributable to the backlash of the mechanical rotator used for the experiments, i.e., $\Delta\theta \approx 1^{\circ}$ and $\Delta\phi \approx 1^{\circ}$. The important point is that the power law exponent of the magnetoresistivity remains at a value $n=1$, or well below it, as the field perpendicular to the current increases, displaying a quadratic term only at the lowest fields.  Below we argue that sublinear behavior is expected for a non carrier compensated system with both types of carriers having distinct mobilities. In contrast, the linear behavior for the LMR can result from the chiral anomaly among type-II Dirac nodes. Notice, that this unconventional magnetotransport behavior, precludes a reliable extraction of carrier densities and mobilities. This would require a simultaneous fitting of both the Hall-effect and the MR to semiclassical equations, which cannot describe the linear magnetoresistivity over  the entire field range.

\subsection{Band structure calculations and Fermi surface  through the de Haas-van Alphen effect}

We proceed to evaluate the geometry of the Fermi surface (FS) of Pd$_3$In$_7$ through the de Haas-van Alphen (dHvA) effect, in an attempt to correlate its geometry with the one predicted by band structure calculations.
The magnetic torque, $\mathbf{\tau} = V \mathbf{M} \times (\mu_0 \mathbf{H}) $, where $V$ is the volume of the sample and $M$ its magnetization, is shown in Fig. 4a as a function of the magnetic field. The sample was rotated according to the sketch in Fig. 4 starting with $\mu_0\textbf{H} \parallel (0,-1,1)$ corresponding to $\theta'=0^{\circ}$ and ending with $\mu_0\textbf{H} \parallel (0,1,1)$ or $\theta'=90^{\circ}$. This particular plane of rotation was defined by the geometry of the sample given that its (0,-1,1) plane was the largest. The crystal was first characterized via a 4-probe resistivity measurement, before it was cut for the magnetic torque magnetometry. As seen in Fig. 4a, one observes oscillations in $\tau$ at $T = 0.35$ K due to the dHvA effect, starting at approximately 2 – 3 T, which implies carrier mobilities exceeding 3000 cm$^2$ V$^{-1}$s$^{-1}$. It turns out that this value is  smaller than $\mu_{\text{tr}}$, but it could result from the inherent lack of sensitivity of the torque technique at the lowest fields. The oscillatory component, or the dHvA signal, which is obtained after subtracting the background through a polynomial fit, is plotted as a function of inverse magnetic field $(\mu_0 H)^{-1}$ in Fig. 4b.
Figures 4c and 4d display the Fast Fourier transform (FFT) of the dHvA signal to extract the superimposed frequencies $F$. Many frequencies, and their harmonics, can be observed in the range 0 to 4 kT, which correspond to extremal cross-sectional areas of the different FS sheets, according to the Onsager relation, $F = \hbar /2\pi e \cdot A$, where $A$ is the cross-sectional area, $\hbar$ is the reduced Planck constant and $e$ the electron charge. The magnitude of selected frequencies as a function of the temperature are shown in Supplementary Fig. 3 in the SI file. The experimental data have been fitted using the temperature damping factor of the Lifshitz-Kosevich formalism, i.e.,  $X/\sinh(X)$, where $X = 2\pi^2 k_\text{B} Tm^{\star}/\text{e} \hbar \mu_0H$, $m^{\star}$ is the carrier effective mass in units of the free electron mass $m_e$ \cite{shoenberg_1984}. Supplementary Table 2 summarizes all the detected frequencies and their corresponding effective masses. As seen, all the experimentally obtained effective masses display values within the $\sim 0.08 - 0.24$ $m_{\text{e}}$, with those associated to the lower frequencies agreeing well with the theoretically predicted values. On the other hand, the theoretical band  masses $m_{\text{b}}$ associated with the higher frequencies are overestimated in comparison to the measured effective masses, indicating that the electronic bands yielding these FS sheets disperse more linearly than predicted by the calculations.


Interestingly, the amplitude of these dHvA frequencies is markedly angle dependent. For instance, the $\kappa$-branch reveals the existence of the so-called spin-zeros or angles where the magnitude of the oscillations vanishes due to the spin-dephasing factor in the Lifshitz-Kosevich formalism, i.e.,  $R_{\text{s}} = \cos(\pi gm^{\star}/2m_0)$, where $g$ is the Land\'e $g$-factor \cite{shoenberg_1984}. This term reaches zero whenever $\pi gm^{\star}/2m_0 = (2n+1)\pi /2$. Supplementary Fig. 7, displays the FFT magnitude of the  $\kappa$-branch as a function of $\theta^{\prime}$, revealing two spin-zeros from which one can estimate both the value and the anisotropy of the $g$-factor. To calculate the $g$-factor, we use two approaches, one based on DFT calculations and the other on experimental results. According to the theoretically calculated effective masses shown in Supplementary Fig. 8, at the first spin-zero occurring at $\theta^{\prime}\sim 30^{\circ}$, chosen by us to correspond to $n=0$, one obtains $g \approx 1.96$. For the second at $\theta^{\prime}\sim 80^{\circ}$, or $n=1$, one obtains $g \approx 5.26$. From Supplementary Fig. 8 it is evident that the $m_{\kappa}^{\star}$ associated to the $\kappa$-branch does not change considerably as a function of $\theta^{\prime}$. In fact, even if we did not choose specific values for  $m_{\kappa}^{\star}$, the values of $g$ for the first spin-zero would fall within $1.64 \leq g \leq 2.04$, and $4.92 \leq g \leq 6.12$ for the second. The other approach is to use the experimentally obtained effective masses. In Supplementary Fig. 9 we provide values of $m_{\kappa}^{\star}$ for two angles, i.e., $\theta^{\prime} = 0^{\circ}$ and $40^{\circ}$. The DFT calculations in Supplementary Fig. 8 imply that $m^{\star}$ does not change much within the range $\theta^{\prime}=30^{\circ}-60^{\circ}$. Therefore, the experimental value of $m^{\star}$ at $\theta^{\prime}=40^{\circ}$ can be safely used to estimate $g$ at the first spin-zero, or $n=0$, yielding $g=9.1$. For the second spin-zero, or $n=1$, the value of $m_{\kappa}^{\star}$ at $\theta^{\prime}=0^{\circ}$ can be used, since our calculations indicate this value to be equal to the value of $m_{\kappa}^{\star}$ at $\theta^{\prime}=90^{\circ}$. This effective mass value would yield $g=17.64$. Regardless of the precise values used, one can safely conclude that the Land\'{e} $g$-factor in Pd$_3$In$_7$ is anisotropic and displays large values. Values that are considerably larger than the usual $g= 2$ value assumed for free electrons, pointing to a pronounced orbital contribution to $g$. The immediate consequence of a large and anisotropic $g$-factor, would be a very pronounced and anisotropic Zeeman-effect, capable of spin polarizing and therefore deforming the geometry of the smaller Fermi surfaces of Pd$_3$In$_7$. This, in turn, is likely to affect the scattering rates and hence carrier mobilities on the FS sheets.

Now we proceed to compare the geometry of the experimentally determined FS sheets with those resulting from our electronic band structure calculations. A good agreement between them would support the validity of the calculations, and hence the existence of Dirac nodes in Pd$_3$In$_7$. Fig. 5a displays the first Brillouin zone of Pd$_3$In$_7$ along with its high symmetry points as well as high symmetry directions connecting them. The actual electronic band structure, along the main reciprocal lattice directions of Pd$_3$In$_7$, is shown in Figs. 5b and Supplementary Fig. 11. Blue lines depict calculations including spin-orbit coupling (SOC) while the orange ones exclude SOC. As seen in Figs. 5b, 5c, and Supplementary Fig. 11, one  easily identify a total of five tilted Dirac type-II nodes along the $\Gamma-$H, $\Gamma-$N, and P-H directions that are within $\pm 250$ meV from the Fermi level $\varepsilon_{\text{F}}$, with some of the associated bands dispersing linearly all the way up (or down) to $\varepsilon_{\text{F}}$. In Fig. 5c, the crossings leading to type-II Dirac nodes are indicated by green markers. Notice that several of the crossings observed in Fig. 5b become gaped by the SOC, as is illustrated in Fig. 5c by the red dashed line encircling two nearly touching bands. Other crossings lead to Dirac nodal lines (see, Supplementary Fig. 11). To illustrate the presence of multiple type-II Dirac nodes in this system, we show in Fig. 5d a three-dimensional plot of a specific crossing, or a type-II Dirac band touching, labelled as crossing 1 in Fig. 5c, which is relatively close to $\varepsilon_{\text{F}}$. See methods for calculation details.

Experimentally, the topography of the FS can be mapped out via the angular dependence of the dHvA frequencies. A good agreement between our observations and the theoretically predicted angle-dependent FS cross-sectional areas would support our calculations and therefore the existence of Dirac nodes close to  $\varepsilon_{\text{F}}$ in Pd$_3$In$_7$. Here, the fundamental question is if there is a correlation between the type-II Dirac nodes and the anomalous magnetoresistivity displayed by Pd$_3$In$_7$. But first, we must confirm their presence by comparing calculated and experimentally determined FS geometries.

Colored panels, in Fig. 5e, display the FS sheets for the different bands intersecting $\varepsilon_{\text{F}}$. In Supplementary Fig. 10 the electronic band structure is displayed with the bands intersecting  $\varepsilon_{\text{F}}$ having different line colors to convey the respective bands that yield each of the FS sheets shown in Fig. 5e. While grey markers in Fig. 5f depict the angular dependence of the observed dHvA frequencies $F$, colored markers correspond to the theoretically calculated FS cross-sectional areas (in Fig. 5e). Bands 1 through 3 yield hole pockets, while bands 4 and 5 produce electron-like sheets. In Fig. 5f, there is a good agreement between the experimental and theoretical points for the high frequencies, or FS sheets, from bands 2 and 3. These hole pockets are the FSs with the largest volume relative to the volume of the Brillouin zone, and therefore they contribute the most to the density of carriers. For the smaller volume FS sheets (or lower frequencies), there are discrepancies between the theoretical and experimental points. In our experience, minor discrepancies between FSs determined through quantum oscillatory phenomena and those resulting from Density Functional Theory (DFT) calculations are rather common, see for example Refs. \cite{Klotz, Shirin}. One can understand this from the fact that different DFT implementations, for example using distinct electron-density functionals (GGA, PBE, HSE06, etc.), tend to yield slightly different positions for the  electron and hole bands relative to $\varepsilon_{\text{F}}$. Therefore, we estimate the typical DFT error bar to be  in the order of few tens of meV, by comparing the distinct DFT implementations. However, such small energy value, or small displacement relative to $\varepsilon_{\text{F}}$, can  strongly impact the sizes of the smaller FS sheets.

 Additionally, multiple frequencies, harmonics and combinations of harmonics are observed due to the complexity of the calculated Fermi surface, and the high quality of the single crystals of Pd$_3$In$_7$ (see Figs. 4c and 4d). This makes it challenging to distinguish fundamental frequencies from harmonics and their combinations. Therefore, we conclude that Fig. 5 points to a quite honest agreement between calculations and the experimental data, thus supporting the existence of Dirac type-II nodes in Pd$_3$In$_7$. This conclusion is further supported by the small effective masses tabulated in the Supplementary Table 2, albeit they are not perfectly captured by the calculations that yield heavier masses for the larger Fermi surface sheets. This indicates that DFT overestimates the curvature of some of the bands in the vicinity $\varepsilon_{\text{F}}$. The small  experimental quasiparticle effective masses, for all FS sheets, further support the notion of Pd$_3$In$_7$ being a Dirac like compound. Despite minor disagreements with the experiments, DFT reveals the presence of multiple type-II Dirac nodes in Pd$_3$In$_7$ near $\varepsilon_{\text{F}}$ and this is crucial for interpreting our observations, as we outline below.


\section{\label{sec:level3}DISCUSSION}
From the magnetoresistivity measurements we extracted four important observations: i) it displays a linear in field behavior from $\mu_0H = 0$ T all the way up to the highest  field value of $\mu_0H = 41$ T  when $\theta = 0^{\circ}$, ii) for all other values of $\theta$, the magnetoresistivity displays a superposition of linear plus quadratic in field terms that becomes sub-linear when $\mu_0H$ exceeds 4 T, with a power exponent as low as $n = 0.7$, iii) for a metal characterized by large Fermi surface sheets (or, in contrast to nearly compensated semimetals), the magnetoresistivity is very large, reaching $10^4$ $\%$  under $\mu_0H = 41$ T, and iv) its anisotropy leads to a butterfly like angular dependence that was previously reported for Dirac nodal line semimetals.

Linear CMR starting from very low magnetic fields was  previously reported for the pnictide FeP \cite{FeP}, where a combination of topological band structure with helimagnetic order, led the authors to propose intra-band scattering among topological bands for its origin. Linear CMR was also observed in high mobility semiconducting GaAs quantum wells \cite{Khouri} and ascribed to the so-called resistance rule, dictating that their magnetoresistivity should be proportional to the derivative of the Hall response with respect to the induction field $B$ or $R_{\text{xx}} = dR_{\text{xy}}/dB \times B \times \alpha$, where $\alpha$ is a constant of proportionality. However, and as we show in Supplementary Fig. 12, $dR_{\text{xy}}/dB \times B$ is not at all linear for Pd$_3$In$_7$, indicating the  irrelevance of this scenario for this compound. Linear CMR in the cuprates \cite{Shekhter1}, and also in the Fe based superconductors \cite{Hayes}, was attributed to the field-induced renormalization of the dynamics intrinsic to their quantum critical fluctuations (associated to suppression of their magnetic order) and hence of their concomitant quasiparticle lifetime. However, Pd$_3$In$_7$ is non-magnetic, does not exhibit long-range charge-order, and does not reveal evidence for pronounced electronic correlations due to a possible proximity to a quantum critical point. Therefore, these previously proposed mechanisms are unable to address its linear magnetoresistivity, which in contrast to these examples, is observed also for fields aligned along the electrical currents, or in a LMR configuration. Linear CMR was reported for Dirac and Weyl semimetals, like Cd$_3$As$_2$ \cite{WOS:000350136400013,PhysRevLett.114.117201}, NbP \cite{NbP}, TiBiSSe \cite{TiBiSSe},  graphene \cite{epitaxial-graphene,bilayer-graphene}, topological insulators like YPdBi \cite{YPdBi} or Bi$_2$Te$_3$ nanosheets \cite{Bi2Te3}, as well as charge/spin density-wave materials \cite{density-wave}.  Given the observation of linear CMR in multiple Dirac/Weyl systems including Pd$_3$In$_7$, see Figs. 3c and 3d, one is led to conclude that linear in field magnetoresistivity is intrinsic to linearly dispersing electronic bands. Notice that linear LMR  was previously reported for WTe$_2$ \cite{PhysRevB.92.041104}, a compound also proposed to be a type-II Weyl semimetal \cite{Bernevig}. Although, in contrast to Pd$_3$In$_7$, WTe$_2$ displays a quadratic in field CMR \cite{WTe2}.

The theory of the quantum linear magnetoresistivity by Abrikosov \cite{Abrikosov}, cannot be applied to Pd$_3$In$_7$. According to it, a very small Fermi surface pocket dominates the magnetoresistivity when it is driven to the quantum limit by an increasing field, or when all carriers fill the lowest Landau level. In our case, the observed as well as the calculated de Haas-van Alphen frequencies, exceed $\mu_0H = 100$ T, implying that during the measurements we remained far below the quantum limit for all Fermi surface sheets.  In Cd$_3$As$_2$ the linear MR is attributed to the magnetic-field induced lifting of a proposed topological protection against backscattering  \cite{WOS:000350136400013, PhysRevB.92.081306} or to mobility fluctuations caused by disorder in this system \cite{PhysRevLett.114.117201}. We have no experimental evidence pointing to topological protection in Pd$_3$In$_7$. As for mobility fluctuations, it is very unlikely that such scenario would also explain its sublinear magnetoresistivity as the crystal is tilted, or the angular dependence of the power law exponent $n$. Instead, we discuss below the possible role of a large anisotropic Zeeman-effect on the geometry of the Fermi surface of Pd$_3$In$_7$ which we postulate to be a non-compensated metallic system. Notice that we cannot extract its carrier densities through fittings of our transport data to conventional transport models because these assume a quadratic in field behavior for the magnetoresistivity.

Pd$_3$In$_7$ has multiple type-II Dirac nodes (Figs. 5b, 5c, and Supplementary Fig. 11) dispersing along distinct directions within its Brillouin zone. Under a magnetic field, each type-II Dirac cone splits into type-II Weyl cones of opposite chirality due to the Zeeman effect. The vector connecting them is defined by both the Zeeman-effect and the spin-orbit coupling (SOC) acting on each band. Now, as the sample is rotated, the high Land\'{e} $g$-factor of the system can trigger a strong and anisotropic Zeeman response, modifying the geometry and hence the scattering rate on the FSs. As a result, the ratio between electron  and hole mobilities $(\mu_{\text{e}}/\mu_{\text{h}})$ is likely to vary with the field. Now, as we show below, varying  $\mu_{\text{e}}/\mu_{\text{h}}$ for a system with uncompensated carrier density ($n_{\text{e}}/n_{\text{h}}\neq 1$) explains the linear as well as the sublinear dependence of the power-law exponent $n(\theta)$ of the CMR, even when these ratios are assumed to be field independent.

To better emphasize our point, we consider a simple two-band model, for which the semiclassical conductivity  is given by the following expression in complex representation~\cite{WTe2}
\begin{align}
{\sigma}=e\left[\frac{n_{\text{e}} \mu_{\text{e}}}{(1+i \mu_{\text{e}} \mu_0H)}+\frac{n_{\text{h}} \mu_{\text{h}}}{\left(1-i \mu_{\text{h}} \mu_0H\right)}\right],
\label{Eq:sigmnatensor}
\end{align}
where $n_{\text{e}}$ and $n_{\text{h}}$ are the carrier densities of electrons and holes, respectively, and $\mu_{\text{e}}$ and $\mu_{\text{h}}$ are their corresponding mobilities. Eq.~\ref{Eq:sigmnatensor} can be inverted to find the resistivity $\rho$ in complex representation. The real part of $\rho$ provides  $\rho_{\text{xx}}$.

In Fig.~\ref{fig:powerlaw_theory} the exponent $n$, corresponding to the power-law, $\rho_{\text{xx}} \sim (\mu_0 H)^n$ is plotted for the above two-band model. For illustrative purposes, in the color contour plot of $n$ displayed in Fig.~\ref{fig:powerlaw_theory}a, we assume that the electron and hole carriers are uncompensated by fixing $n_{\text{h}}/n_{\text{e}}=0.75$, while allowing both the magnetic field and  the mobility ratio $\mu_{\text{h}}/\mu_{\text{e}}$ to vary. In Fig.~\ref{fig:powerlaw_theory}b, we display a few traces leading to the color contour plot.
The ratio $n_{\text{h}}/n_{\text{e}}=0.75$ is not based on experimental values extracted from measurements on Pd$_3$In$_7$ since, as previously mentioned, one cannot satisfactorily fit its transport data to conventional transport models. In fact, when we attempted to fit the Hall conductivity to the two-band model, it tended to converge to very large, non-physical ratios for the electron and hole mobilities, and respective densities, suggesting that our interpretation is correct. It turns out that from Eq.~\ref{Eq:sigmnatensor}  the behavior of the exponent $n$ with respect to the magnetic field and the mobility ratio $\mu_{\text{h}}/\mu_{\text{e}}$ is in general non-monotonic. The important observation is that the exponent $n$ can fall below two and all the way to one (linear dependence) and even below one (sublinear dependence). In Figs.~\ref{fig:powerlaw_theory}c and~\ref{fig:powerlaw_theory}d we maintained the mobility ratio fixed at $\mu_{\text{h}}/\mu_{\text{e}} = 0.5$ but varied the ratio $n_{\text{h}}/n_{\text{e}}$. As seen, the magnetoresistivity  displays a sublinear dependence on the external field for all values of $n_{\text{h}}/n_{\text{e}}$. Experimentally, we determine that $1/\mu_{\text{e}} \approx \mu_0 H_0 \approx 0.8$ T (Fig. 1d) corresponding to the magnetic field range from 0 to 32 T in Fig.~\ref{fig:powerlaw_theory}.

Since varying the angle $\theta$ in Fig.~1(b) can change the mobility ratio $\mu_{\text{h}}/\mu_{\text{e}}$ ($k$-dependent scattering rate on the Fermi surfaces), the angular behavior of the exponent $n(\theta)$ in Fig.~3(c) is explained.
We do not have an explanation for the evolution of the power law prefactor $\alpha$. We point out that sublinear magnetoresistivity was also detected in the type-II Dirac semimetal NiTe$_2$ \cite{NiTe2}.


As seen through Figs. 5 and Supplementary Fig. 11, the spin orbit coupling gaps a series of linear band crossings, but it still leaves others ungapped. The DFT calculations indicate that there are two non-gapped type-II Dirac  nodes along the $\Gamma-$H direction, a third one along the $\Gamma-$N direction, and a fourth one along the P$-$H direction, all located within $\sim$ 250 meV of $\varepsilon_{\text{F}}$. The calculations indicate that other bands, not associated to linear band crossings or Dirac like nodes, and therefore topologically trivial, also intersect $\varepsilon_{\text{F}}$. However, as previously discussed, there are no known mechanisms that would explain the linear LMR based on topologically trivial bands and associated Fermi surfaces. In contrast, and as we discuss below, there are published predictions correlating the linear LMR with the chiral anomaly among type-II Dirac  nodes \cite{Kondo,knoll2020negative,ahmad2021longitudinal}. Given the absence of an alternative explanation, and the existence in a very clean system like Pd$_3$In$_7$ of several Dirac type-II like nodes near $\varepsilon_{\text{F}}$, the only possible conclusion is to attribute the linear LMR to the non-trivial topology of the associated bands, despite  not being located exactly at $\varepsilon_{\text{F}}$.

In effect, when the magnetic field has a nonzero component in the direction of the electric field, as in Fig. 1a, the linear and quadratic dependence on the magnetic field, as well as the sign of the positive and linear LMR can  be understood purely due to chiral anomaly~\cite{Kondo,knoll2020negative,ahmad2021longitudinal}. It was initially anticipated that the chiral anomaly contribution to the LMR is always negative and quadratic. 
However, it was shown that the magnetoresistivity can be linear as well as positive for sufficient tilting of the Weyl cones in the direction of the magnetic field~\cite{ahmad2021longitudinal,Kondo}. The change in the sign of the magnetoresistance, i.e., from negative to positive, results from a finite intervalley scattering. In the Supplementary Information, we explicitly show the emergence of positive linear magnetoresistance for a model of tilted Weyl fermions, see, Supplementary Discussion and Supplementary Fig. 13. Even though our calculation is restricted to Weyl cones with the tilting parameter less than unity, we expect the qualitative features to remain the same even for larger values of the tilt, as also shown in earlier works~\cite{sharma2017chiral}.  Here, we emphasize that for the LMR configuration, there is no Lorentz force acting on the charge carriers. Therefore, one cannot calculate the LMR using the semiclassical approach of Eq. 3 that yielded a power law inferior to 1 for the CMR.
In the present case, the natural mechanism that would yield linear longitudinal magneto-resistivity, as observed by us, is the chiral anomaly among type-II tilted Weyl nodes in broad agreement with Refs. \cite{sharma2017chiral,ahmad2021longitudinal,Kondo}. However, a precise comparison with these models would require additional DFT calculations that consider the Zeeman-effect as well as the spin-orbit coupling. The goal would be to determine the precise $k$-space orientation of the vector(s) connecting the field-induced type-II Weyl nodes, and their relative orientation with respect to electric and magnetic fields.

It is frequently assumed that the Dirac/Weyl nodes should be located at, or in the vicinity of $\varepsilon_{\text{F}}$, for the axial anomaly to affect  transport properties. However, this critical distance to $\varepsilon_{\text{F}}$ is system specific, and hence band-structure dependent. This mechanism depends on the remnant  texture of the Berry phase at $\varepsilon_{\text{F}}$, whose sources/sinks are the Dirac/Weyl nodes, and on the exact scattering mechanism(s) at $\varepsilon_{\text{F}}$ within a given system. Without detailed calculations, that are beyond the scope of this manuscript, there is no reason to assume that a distance $\Delta E \simeq 250$ meV, with respect to $\varepsilon_{\text{F}}$, for the type-II Dirac nodes of Pd$_3$In$_7$, should preclude a role for the chiral anomaly. Furthermore, in lattice models for Weyl semimetallic systems with Weyl nodes located away from $\varepsilon_{\text{F}}$, several models predicted a pronounced role for the axial anomaly, see for example, Ref. \cite{Sharma2}.
As for the linear magnetoresistivity  observed in Fig. 3d, over the entire range of the angle $\phi$, one must consider that the axial anomaly is predicted to play a predominant role even when a very small component of the electric field is aligned along the magnetic field \cite{Sharma2, Sharma3}. These models consider not only the role of the inter-valley scattering and the tilting of the cones, but also the non-linearity of the electronic bands away from the Dirac/Weyl nodes \cite{Sharma2, Sharma3}. The angular range in Fig. 3d encompasses two behavioral extremes ranging from linear CMR ($\phi = 0^{\circ}$) to linear LMR ($\phi = 90^{\circ}$). Therefore, it is not surprising that the superposition of both contributions still leads to linear magnetoresistivity at intermediary angles.

We conclude by stating  that our observations indicate that two mechanisms affect the magneto-transport of Pd$_3$In$_7$, i.e., the i) axial anomaly among field-induced type-II Weyl nodes, and ii) a multi-band effect arising from an uncompensated metal subjected to a pronounced, and angle-dependent, scattering rate on its Fermi surface. The observation of linear LMR in two crystallographic distinct compounds, i.e., cubic Pd$_3$In$_7$ and orthorhombic WTe$_2$ \cite{PhysRevB.92.041104}, indicates that it is intrinsically associated to the type-II Dirac nodes in their electronic dispersion. Anisotropy in the scattering rate might result from a large and anisotropic Zeeman effect, or, the large and anisotropic Land\'{e} $g$-factor of Pd$_3$In$_7$. Or, it might be intrinsic to systems lacking Lorentz invariance, as is the case for type-II Weyl/Dirac systems, implying that our observations might be generic and extendable to many other systems. Even though linear LMR and sublinear CMR, have been previously and independently observed in Weyl semimetals, their simultaneous observation in a single system is the distinct hallmark of this work.

\section{\label{sec:level4}METHODS}

\textbf{Sample synthesis}
High quality single crystals of Pd$_3$In$_7$ were synthesized via an In-flux method, where elemental Pd and In, with an atomic ratio 14:86, were loaded in a Al$_2$O$_3$ crucible and sealed in an evacuated fused silica ampule. The tube was heated to 750 $^{\circ}$C and held there for 48 h. Then, it was cooled to 600 $^{\circ}$C at a rate of 4 $^{\circ}$C/h and afterwards slowly cooled to 400 $^{\circ}$C at a rate of 0.5 $^{\circ}$C/h. At this point, the tube was centrifuged and the as-harvested single crystals were etched in diluted HCl to remove residual metal from their surface. The result was the synthesis of shiny crystals with dimensions up to 2 mm.

\textbf{Sample characterization}
The chemical composition of the crystals was determined using single-crystal x-ray diffraction (SCXRD) spectroscopy. Face indexing of the single crystal was performed at room-temperature  using a Rigaku-Oxford Diffraction Synergy-S single-crystal diffractometer equipped with a HyPix detector and a monochromated Mo-K$\alpha$ radiation source ($\lambda$ = 0.71073 \AA). The data set was recorded as a series of $\omega$-scans at $0.5^{\circ}$ step width. The unit cell determination and face indexing were performed with the CrysAlis software package \cite{Crysalis}.

\textbf{Magnetotransport measurements}
For the magnetotransport measurements a Physical Property Measurement System (PPMS) in combination with a rotating probe was used under magnetic fields up to $\mu_0H$ = 9 T and temperatures as low as $T=1.8$ K to perform a four-terminal resistivity measurements. Additional magnetotransport measurements, as well as the angular dependence of the dHvA effect using a piezoresitive microcantilever, were performed under continuous magnetic fields up to $\mu_0H$ = 41.5 T in a resistive Bitter magnet at the National High Magnetic Field Laboratory. The magnet was coupled to a $^3$He cryostat allowing us to perform measurements at temperatures as low as $T = 0.35$ K.

\textbf{Electronic band structure calculations}
The first-principles (DFT) calculations were preformed using the projector augmented-wave (PAW) potentials~\cite{blochl.94} implemented in the Vienna Ab initio Simulation Package ({\sc vasp}) code~\cite{kresse.hafner.94,kresse.furthmuller.96,kresse.joubert.99}.
The calculations were performed in the absence and presence of the spin-orbit coupling (SOC) within the generalized gradient approximation (GGA) developed by Perdew, Burke, and Enzerhof (PBE)~\cite{perdew.burke.96}.
The energy cutoff for the plane-wave expansion is set to $350$~eV.
The electronic DOS was calculated using $15 \times 15 \times 15$ {\bf k}--point grids in the Monkhorst--Pack scheme~\cite{monkhorst.pack.76}.
The Fermi surface was rendered using {\sc XCrysDen}~\cite{kokalj.99}. The frequencies of the Fermi surface were calculated using the Skeaf platform \cite{skeaf}.

\section{\label{sec:level5}Data Availability}
Relevant data supporting the key findings of this study are available within the article and the Supplementary Information file. All raw data generated during the current study are available from the corresponding authors upon request.

\section{\label{sec:level6}Code Availability}
Relevant code for data analysis, data plotting, and band structure calculations are commercially sourced, e.g., the VASP code. Code used to calculated the magnetoresistivity as a function carrier densities and/or mobilities, or to calculate the magnetoresistivity as a function of magnetic field for a Weyl system characterized by tilted Dirac/Weyl  cones, and inter-valley scattering, is available from the authors upon request.


 \section{\label{sec:level7}Acknowledgments}
 We acknowledge very useful discussions with Xiaofeng Qian. L.B. is supported by the US-DoE, Basic Energy Sciences program through award DE-SC0002613.  The National High Magnetic Field Laboratory acknowledges support from the US-NSF Cooperative Agreement Grant number DMR-1644779 and the State of Florida.
A.P. appreciates the funding in the frame of scholarships from the Ministry of Science and Higher Education of Poland for outstanding young scientists (2019 edition, No.~818/STYP/14/2019). The work for the DFT calculations was supported by the National Science Centre (NCN, Poland) under Projects No. 2021/43/B/ST3/02166. G.S. acknowledges support from SERB Grant No. SRG/2020/000134. S.T. acknowledges support from Grant No. NSF 2014157.

\section{\label{sec:level8}AUTHOR CONTRIBUTIONS}
A.F.S and L.B. conceived and led the project. A.F.S. synthesized the samples and performed the transport measurements. A.F.S., B.C. and L.B. performed the high field measurements. A.F.S. analyzed all the experimental data. A.P. performed the theoretical calculations for the electronic structure and the Fermi surface. A.F.S. performed the calculations for the different frequencies of the Fermi surface. J.C., V.L. and M.S. performed the single crystal XRD analysis. G.S. and S.T. proposed the model explaining the linear and sublinear (in magnetic field) conventional magnetoresistivity, as well as the linear longitudinal magnetoresistivity. A.F.S. and L.B. prepared the manuscript with input from all co-authors.

\section{\label{sec:level9}COMPETING INTERESTS}
The authors declare no competing interests.

\section{\label{sec:level10}ADDITIONAL INFORMATION}
\textbf{Supplementary Information} The online version contains supplementary material available at https://doi.org/

\textbf{Correspondence} and requests for materials should addressed to Aikaterini Flessa Savvidou or Luis Balicas.


\begin{figure*}[!htpb]
\centering
\includegraphics[width= \textwidth]{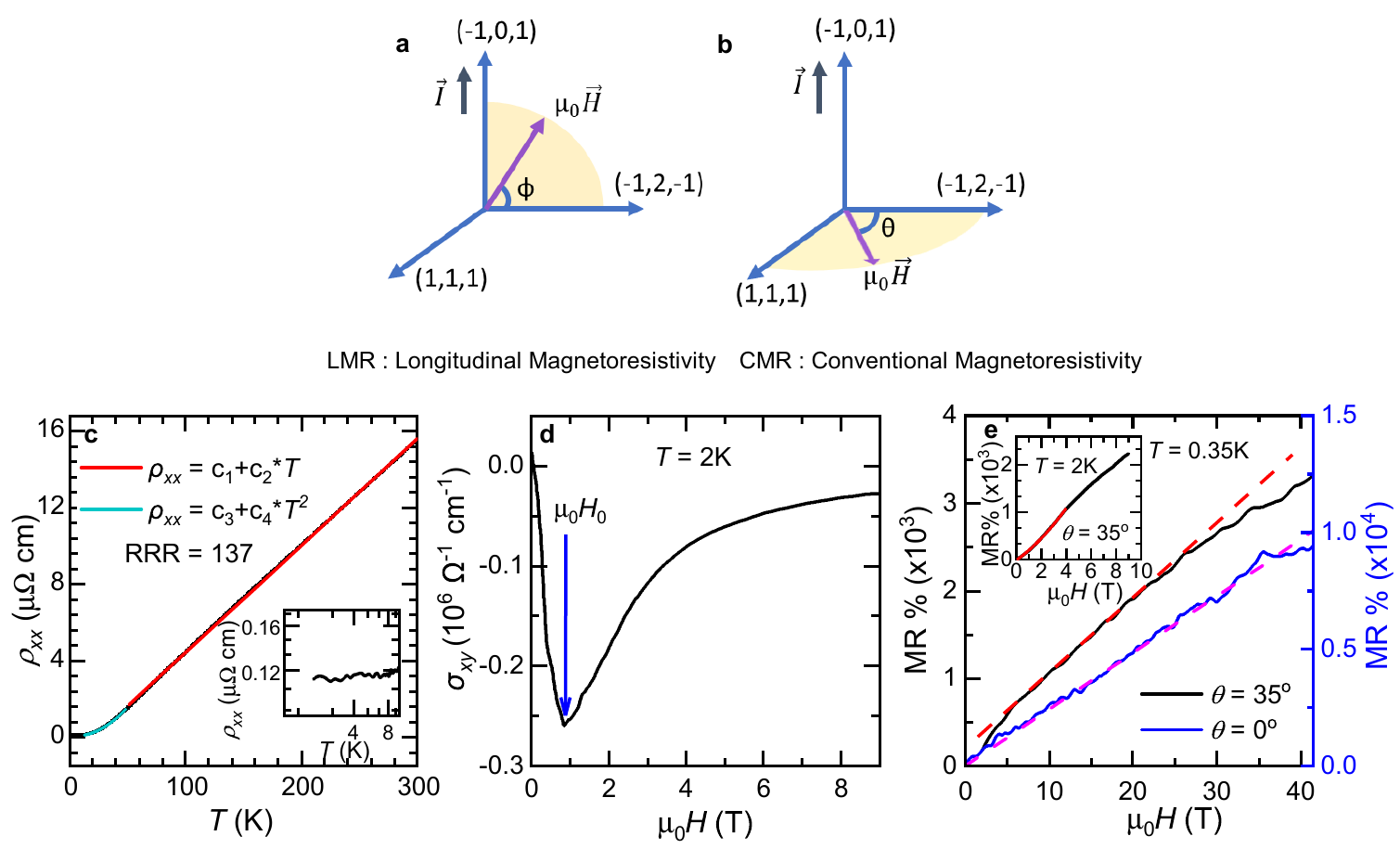}
\caption{\label{fig1:epsart}\textbf{Linear and sub-linear magnetoresistivity in Pd$_3$In$_7$} \textbf{a}, Sketch depicting the planes of rotation of the magnetic field $\mu_0 H$, from $I \perp \mu_0H$ to $I\parallel \mu_0 H$ (longitudinal magnetoresistivity configuration or LMR), as a function of the angle $\phi$. $I$ is the electrical current. \textbf{b}, Sketch illustrating sample rotation under $\mu_0 H$, where $I$ is kept $\perp \mu_0H$ and the angle $\theta$ is varied (conventional magnetoresistivity configuration or CMR) \textbf{c}, Electrical resistivity $\rho_{\text{xx}}$ of Pd$_3$In$_7$ as a function of the temperature $T$. For $T>$ 50 K, $\rho_{\text{xx}}$ displays a linear dependence on $T$ (red  line is a linear fit), while $\rho_{\text{xx}}$ exhibits a quadratic behavior (cyan line is a fit) for $T\leq50$ K. Inset: $\rho_{\text{xx}}$ in a magnified scale illustrating the residual resistivity $\rho_0 \simeq 110$ n$\Omega$ cm. \textbf{d}, Hall conductivity $\sigma_{\text{xy}}$ as a function of $\mu_0 H$, displaying a minimum at $\mu_0 H_0=$ 0.84 T. \textbf{e}, Magnetoresistivity MR as a function of $\mu_0 H$ for 3 T $\leqq$ $\mu_0 H$ $\leqq$  41.5 T.  Blue and black lines correspond to the experimental data for $\theta = 0^{\circ}$ and $\theta = 35^{\circ}$, respectively. Dashed magenta line corresponds to a linear fit of the $\theta = 0^{\circ}$ data for the entire field range, while red dashed line indicates the deviation of the black line with respect to linear behavior. Inset: \emph{MR} as a function of $\mu_0H$, for fields up to $\mu_0H =$ 9 T. Red line is a fit of the data for $\mu_0 H < 4$ T to MR = $d_1$ + $d_2 (\mu_0 H)$ + $d_3 (\mu_0 H)^2$, where $d_2$ and $d_3$ are  coefficients of the linear and quadratic terms, respectively.}
\end{figure*}

\begin{figure*}[!htpb]
\centering
\includegraphics[width=\textwidth]{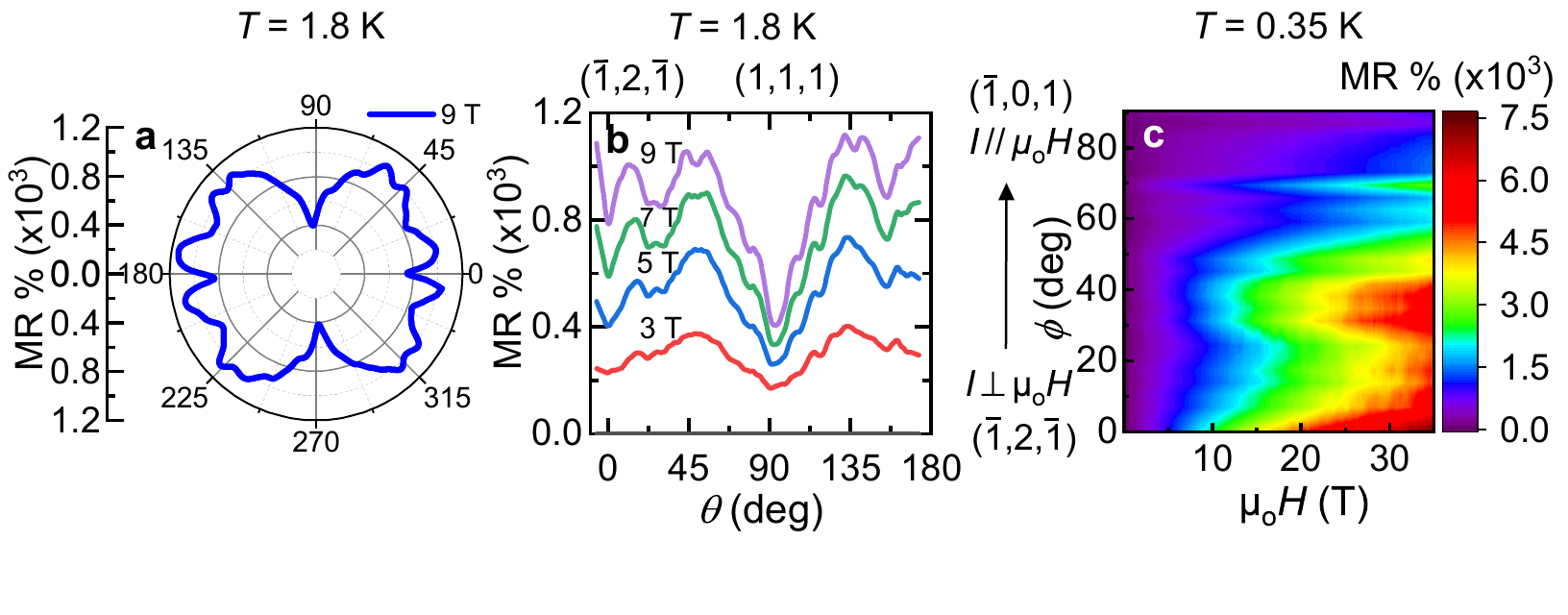}
\caption{\textbf{Butterfly magnetoresistivity in cubic Pd$_3$In$_7$.} \textbf{a}, Polar plot of the magnetoresistivity MR at $T$ = 1.8 K as a function of $\theta$ yielding an anisotropic butterfly-like angular dependence under a magnetic field $\mu_0 H =$ 9 T. $I$ is maintained $\perp$ to $\mu_0 H$ throughout the entire angular range. \textbf{b}, Magnetoresistivity (MR) as a function of $\theta$ from $0^{\circ}$, or $\mu_0\textbf{H} \parallel (-1,2,-1)$ direction, to $\theta = 90^{\circ}$ which corresponds to the $\mu_0\textbf{H} \parallel (1,1,1)$ orientation, for different values of $\mu_0 H$. Throughout the whole rotation at a temperature $T$ = 1.8 K, $I$ is kept $\perp \mu_0H$. The $\Delta \theta = 90^\circ$ periodicity is observed at low fields, i.e., under $\mu_0H = 3$ T, when the amplitude of the Shubnikov de Haas oscillations is quite small, implying that they are not responsible for it. \textbf{c}, Color plot revealing the anisotropy of the MR at $T = 0.35$ K as a function of both angle and field, from $\theta=0^{\circ}$ or $I\perp \mu_0 H$, to 90$^{\circ}$ corresponding to $I\parallel \mu_0 H$.}
\end{figure*}

\begin{figure*}[!htpb]
\begin{center}
\includegraphics[width=0.9\textwidth]{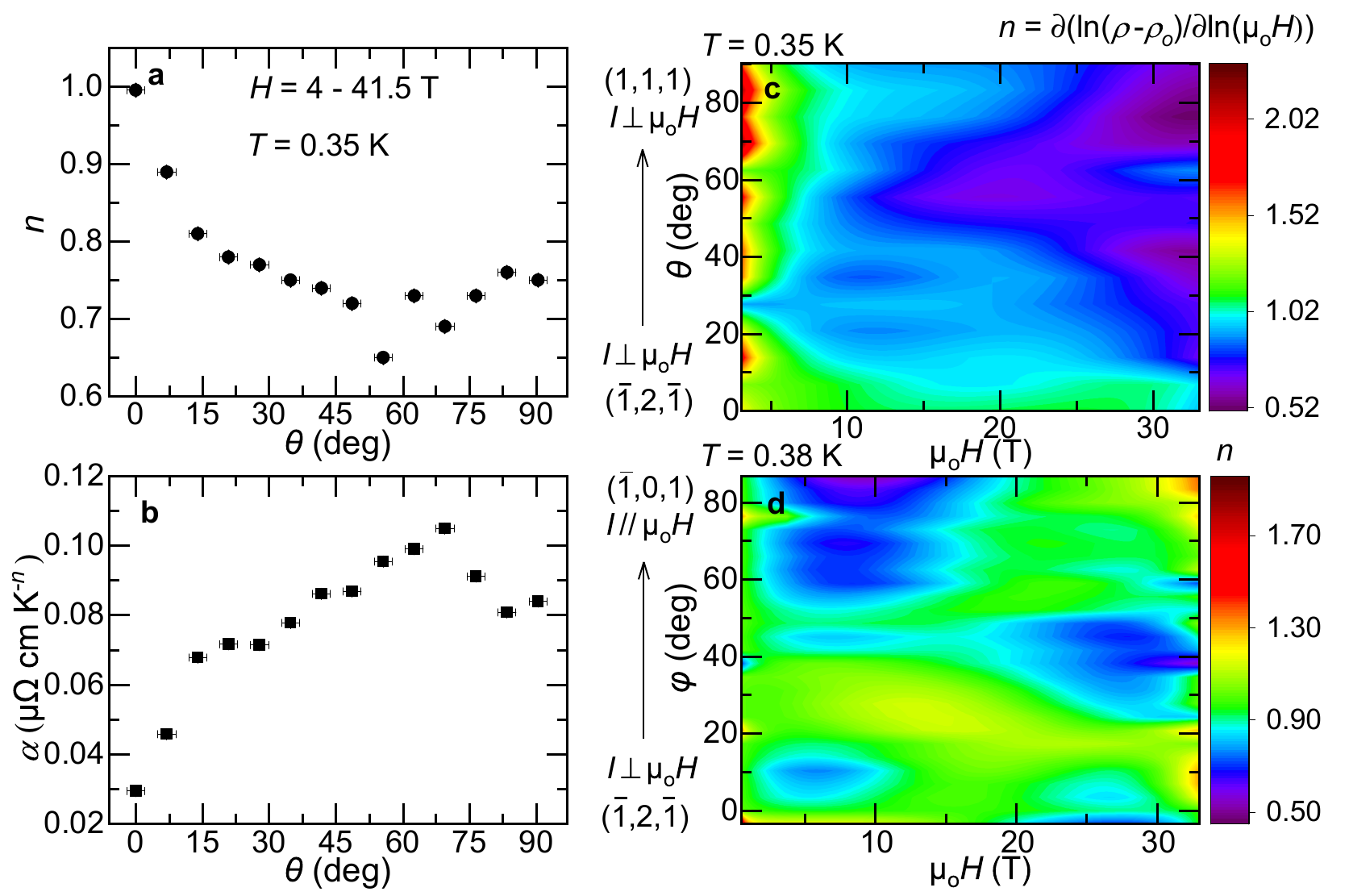}
\caption{\label{fig3:epsart}\textbf{Power law behavior of the magnetoresistivity of Pd$_3$In$_7$.} \textbf{a}, Power law coefficient $n$, in MR $\propto \alpha (\mu_0H)^n$, as a function of the angle $\theta$. Intriguingly, $n$ reveals a minimum around $\theta = 60^{\circ}$. \textbf{b}, Coefficient $\alpha$ as a function of $\theta$. Notice how $\alpha$ increases by a factor $\sim 3$ as $\theta$ exceeds $60^{\circ}$.  \textbf{c}, Color plot of the power coefficient $n$ at $T$ = 0.35 K, extracted from the relation $n=\partial\ln(\rho -\rho_0)/\partial \ln (\mu_0 H)$, as a function of $\theta$ and $\mu_0H$, for $I \perp \mu_0H$ throughout the whole angular range. \textbf{d}, Color plot of the power law coefficient $n$ at $T = 0.38$ K, as a function of $\phi$ and $\mu_0H$, from $\phi = 0^{\circ}$, or $I \perp \mu_0 H$, to $\phi =90^{\circ}$, or $I \parallel \mu_0H$.}
\end{center}
\end{figure*}

\begin{figure*}[!htpb]
\begin{center}
\includegraphics[width=\textwidth]{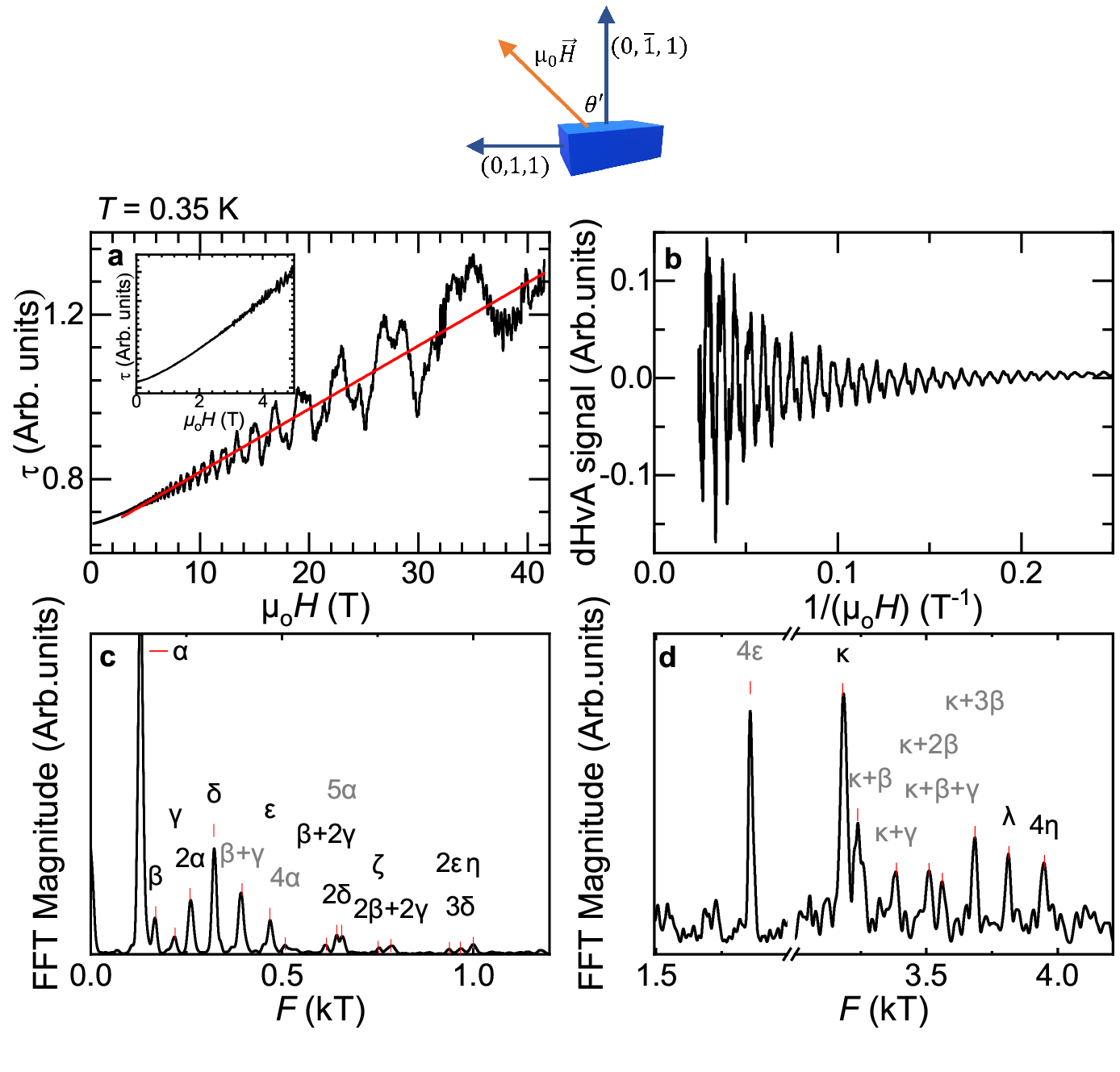}
\caption{\label{fig4:epsart} \textbf{De Haas-van Alphen effect in Pd$_3$In$_7$.} \textbf{a}, Magnetic torque at $T = 0.35$ K and as a function of $\mu_0H$ for the angle $\theta'=0^{\circ}$, where $\theta'$ is the angle between $\mu_0H$ and the $(0,-1,1)$ direction. Red line is a polynomial fit to the background. \textbf{b}, Superimposed de Haas van Alphen (dHvA) signal, as extracted from the magnetic torque in panel a as a function of inverse magnetic field $(\mu_0H)^{-1}$. \textbf{c} to \textbf{d}, Fast Fourier Transform (FFT) of the dHvA signal in b, showing peaks at low and high frequencies that correspond to different extremal cross-sectional areas of the Fermi surface labeled with Greek letters. Peaks labeled by gray colored letters are likely harmonics of the fundamental frequencies, or their combination.}

\end{center}
\end{figure*}

\begin{figure*}[!htpb]
\begin{center}
\includegraphics[width=\textwidth]{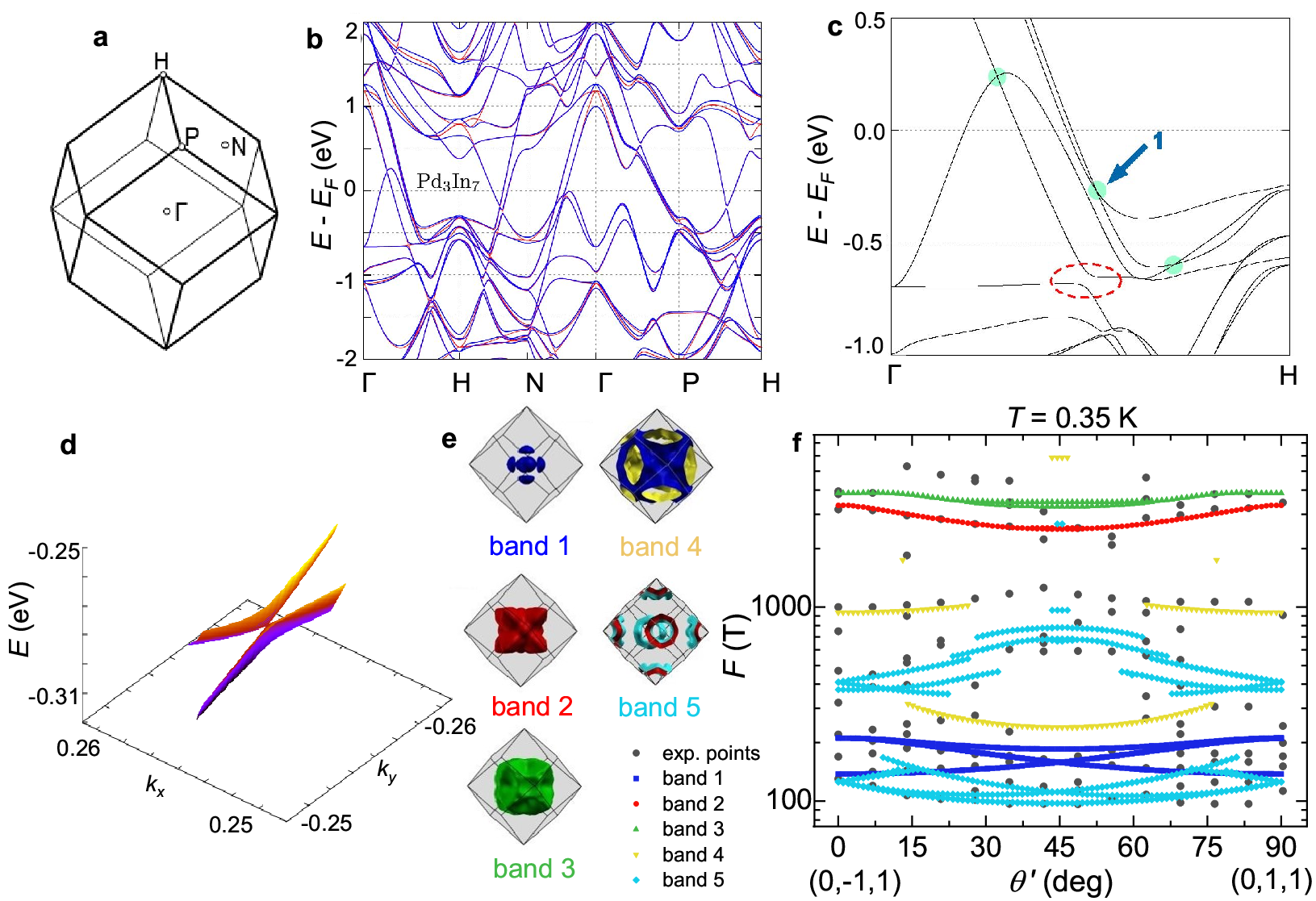}
\caption{\label{fig5:epsart} \textbf{Comparison between calculated and experimental Fermi surface cross-sectional areas.} \textbf{a}, First Brillouin zone (FBZ) of Pd$_3$In$_7$ indicating its high symmetry points and the high symmetry paths explored within its FBZ to evaluate its electronic band structure. \textbf{b}, Theoretically calculated electronic band structure of Pd$_3$In$_7$ in the absence (orange lines) and presence (blue lines) of spin orbit coupling (SOC). \textbf{c}, Electronic band structure of Pd$_3$In$_7$ along the $\Gamma$-H direction, revealing a number of degenerate crossings that lead to type-II Dirac nodes (indicated by green dots). Several crossings become gaped when the SOC is incorporated into the calculations (indicated by dashed red circle). \textbf{d}, Three dimensional plot of the band crossing labelled as 1 in panel c, revealing a type-II Dirac band touching. A few of these crossings were found to produce Dirac nodal lines.
\textbf{e}, Three dimensional plots of the calculated Fermi surface sheets in the FBZ resulting from the distinct bands intersecting the Fermi level.  \textbf{f}, Comparison between the angular dependence of the experimental dHvA frequencies (gray dots) and the theoretically predicted (colored dots) Fermi surface cross-sectional areas of Pd$_3$In$_7$. Here, the colors of the markers were chosen to match those  depicting each FS sheet. See, Supplementary Fig. 10 for a colored depiction of the individual bands leading to each of the plotted Fermi surfaces. }
\end{center}
\end{figure*}

\begin{figure*}[!htpb]
    \centering
 \includegraphics[width=\columnwidth]{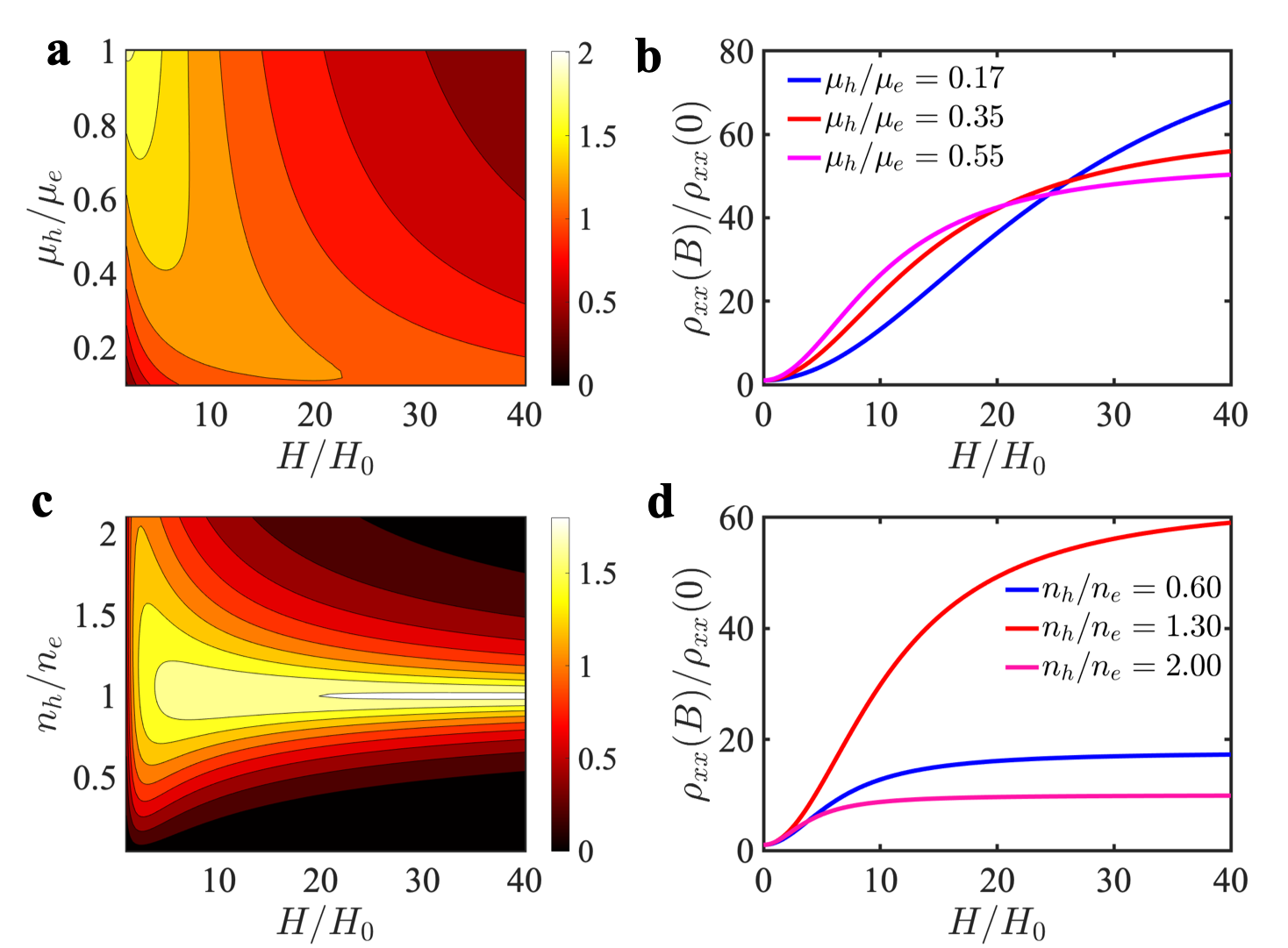}
\caption{\textbf{Origin of the anomalous power law in the conventional magnetoresistivity.} \textbf{a}, Contour plot of the exponent $n$ in the power-law dependence of the magnetoresistivity on the magnetic field ($\mu_0H$), $\rho_{\text{xx}}\sim (\mu_0 H)^n$, evaluated after inverting Eq.~\ref{Eq:sigmnatensor} and for the ratio among carrier mobilities $\mu_{\text{h}}/\mu_{\text{e}}$ ranging from $\sim 0$ to 1. $\mu_{\text{h}}$ and $\mu_{\text{e}}$ stand for hole and electron mobilities, respectively. Here, $\mu_0 H_0 \approx 1/\mu_{\text{e}} \approx 0.8$ T, which corresponds to magnetic fields ranging from 0 to 32 T. \textbf{b}, Calculated resistivity $\rho_{\text{xx}}$ as a function of the external magnetic field for different values of the ratio $\mu_{\text{h}}/\mu_{\text{e}}$. For panels a and b, we chose $n_{\text{h}}/n_{\text{e}} = 0.75$. \textbf{c}, Contour plot of $n$ in the $\mu_0H$ - $n_{\text{h}}/n_{\text{e}}$ plane. It is clear that a marked difference in carrier mobilities, when coupled to even a small imbalance in carrier densities, can favor $n \leq 1$. \textbf{d}, Calculated resistivity $\rho_{\text{xx}}$ as a function of the magnetic field for different values of the ratio $n_{\text{h}}/n_{\text{e}}$. For  panels c and d, we choose $ \mu_{\text{h}}/\mu_{\text{e}} = 0.50$. }
\label{fig:powerlaw_theory}
\end{figure*}

\end{document}


\title[Supplementary Information]{Supplementary information for manuscript:
\\Anisotropic positive linear and sub-linear magnetoresistivity in the cubic type-II Dirac metal Pd$_{3}$In$_{7}$}


\author[1,2]{\fnm{Aikaterini} \sur{Flessa Savvidou}}

\author[3]{\fnm{Andrzej} \sur{Ptok}}
\author[4]{\fnm{G.} \sur{Sharma}}
\author[1]{\fnm{Brian} \sur{Casas}}
\author[5]{\fnm{Judith K.} \sur{Clark}}
\author[5]{\fnm{Victoria M.} \sur{Li}}
\author[5]{\fnm{Michael} \sur{Shatruk}}
\author[6]{\fnm{Sumanta} \sur{Tewari}}
\author*[1,2]{\fnm{Luis} \sur{Balicas}}\email{balicas@magnet.fsu.edu}

\affil[1]{\orgdiv{National High Magnetic Field Laboratory}, \orgname{Florida State University}, \orgaddress{ \city{Tallahassee}, \postcode{32310}, \state{Florida}, \country{USA}}}

\affil[2]{\orgdiv{Department of Physics}, \orgname{Florida State University}, \orgaddress{ \city{Tallahassee}, \postcode{32306}, \state{Florida}, \country{USA}}}

\affil[3]{\orgdiv{Institute of Nuclear Physics}, \orgname{Polish Academy of Sciences}, \orgaddress{\street{W. E. Radzikowskiego 152}, \city{Krak\'{o}w}, \postcode{PL-31342}, \country{Poland}}}

\affil[4]{\orgdiv{School of Physical Sciences}, \orgname{Indian Institute of Technology Mandi}, \orgaddress{ \city{Mandi}, \postcode{175005}, \state{H.P.}, \country{India}}}

\affil[5]{\orgdiv{Department of Chemistry and Biochemistry}, \orgname{Florida State University}, \orgaddress{ \city{Tallahassee}, \postcode{32306}, \state{Florida}, \country{USA}}}

\affil[6]{\orgdiv{Department of Physics and Astronomy}, \orgname{Clemson University}, \orgaddress{ \city{Clemson}, \postcode{29634}, \state{South Carolina}, \country{USA}}}
\date{\today}

\maketitle



\begin{table}[ht]
\renewcommand{\tablename}{Supplementary Table}
\renewcommand{\thetable}{\arabic{table}.}
\centering
\caption{\textbf{List of Supplementary Figures}}

\begin{tabular}{ll}
\toprule
Figure name & Caption \\
\midrule
Supplementary Figure 1          &  Raw magnetoresistivity data for several angles $\theta$.       \\
Supplementary Figure 2          &    Raw magnetoresistivity data for several angles $\phi$.     \\
Supplementary Figure 3          &   Effective masses as extracted from the de Haas-van Alphen signal.      \\
Supplementary Figure 4          &    Sublinear magnetoresistance.     \\
Supplementary Figure 5          &  Kohler plot at $\theta=0^\circ$.       \\
Supplementary Figure 6          &  Magnetoresistance at $\phi=90^\circ$.       \\
Supplementary Figure 7          &  Spin-zeros for the $\kappa$-branch.       \\
Supplementary Figure 8          &    Angular dependence of the effective mass.     \\
Supplementary Figure 9          &   Experimental effective masses for two orientations.      \\
Supplementary Figure 10         &   Electronic structure of Pd$_3$In$_7$.      \\
Supplementary Figure 11         &    Electronic structure between high symmetry points in $k$-space.     \\
Supplementary Figure 12         &   Hall-effect for Pd$_3$In$_7$ and its derivative with respect to field.      \\
Supplementary Figure 13         &    Modeled longitudinal magnetoresistance. \\
\botrule
\end{tabular}
\end{table}

\begin{table}[ht]
\renewcommand{\tablename}{Supplementary Table}
\renewcommand{\thetable}{\arabic{table}.}
\centering
\caption{\textbf{Experimental dHvA results.} dHvA frequencies $F$, their corresponding effective masses $m^{\star}$, with the calculated standard deviation, and calculated band masses $m_b$ for Pd$_3$In$_7$. }
\label{tab:results}

\begin{tabular}{lccc}
\toprule
$F$ (T) & Branch & $m^{\star}$ (exp) ($m_e$) & $m_b$ (theory) ($m_e$) \\
\midrule
130    & $\alpha$    & 0.08 $\pm $ 0.00    & 0.12        \\
170    &$\beta$    & 0.13 $\pm $ 0.01    & 0.10       \\
220    & $\gamma$    & 0.11  $\pm $ 0.01   & 0.16        \\
322    & $\delta$    & 0.15 $\pm $ 0.01    & 0.21        \\
469    & $\epsilon$    & 0.18  $\pm $ 0.01   & 0.23        \\
752    & $\zeta$    & 0.11  $\pm $ 0.02   & ~           \\
1000   & $\eta$    & 0.19 $\pm $ 0.03     & 0.19       \\
3182   & $\kappa$    & 0.17  $\pm $ 0.01     & 0.61       \\
3815   & $\lambda$    & 0.25  $\pm $ 0.03     & 0.56   \\
\botrule

\end{tabular}

\end{table}

\begin{figure*}[!htpb]
\renewcommand{\figurename}{Supplementary Figure}
\renewcommand{\thefigure}{\arabic{figure}.}
\centering
\includegraphics[width=\textwidth]{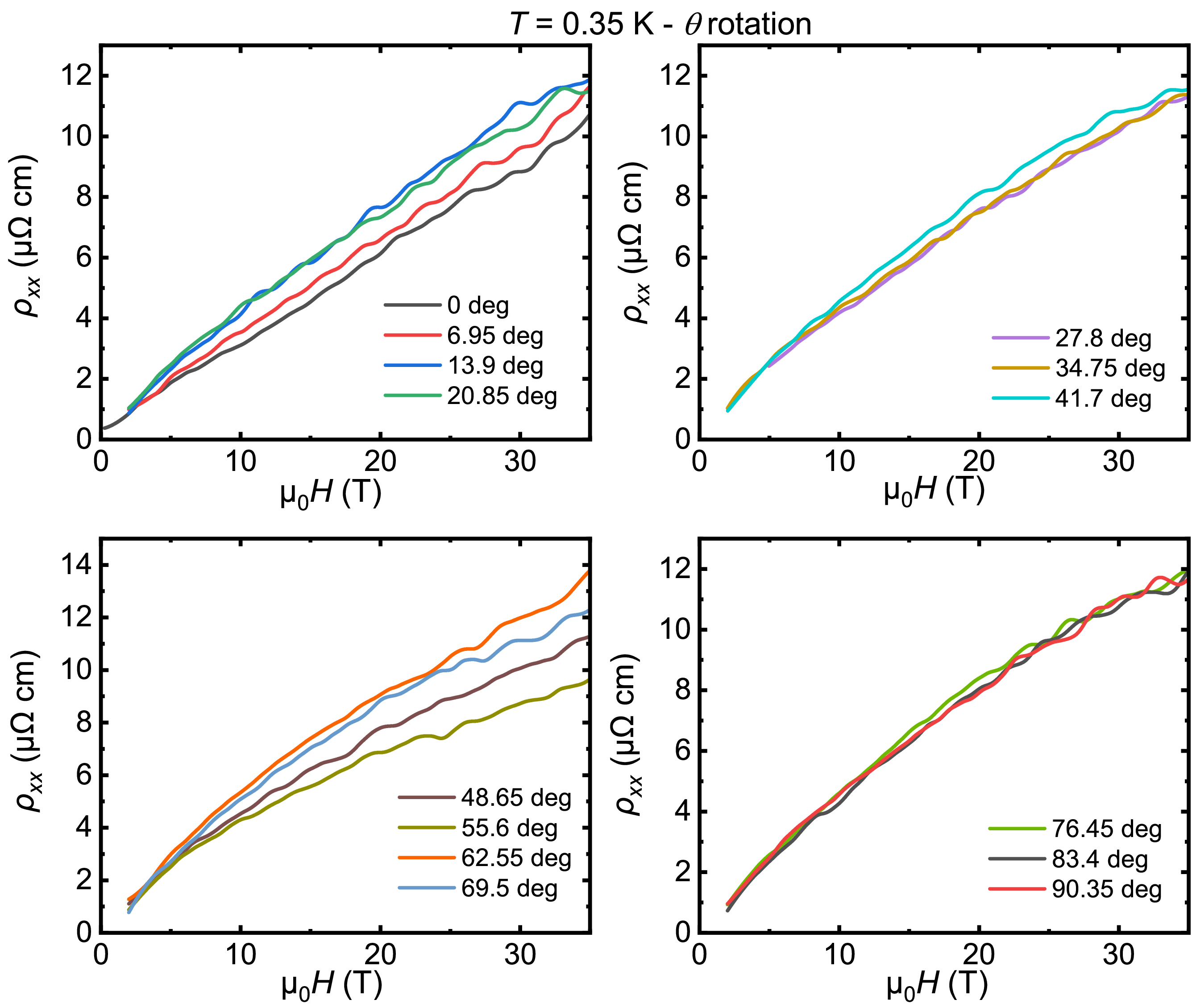}
\caption{\label{fig1:epsart}\textbf{Raw data of the linear and sublinear magnetoresistivity for several angles $\theta$.} Raw magnetoresistivity data as a function of the magnetic field $\mu_0H$ for different rotation angles $\theta$ while keeping $I \perp \mu_0H$ throughout the whole angular range, at a temperature $T$ = 0.35 K. Notice that the $\theta = 0^{\circ}$ traces in top left panel, and  Fig. 1e within the main text, display a linear dependence on $\mu_0H$, albeit with superimposed Shubnikov-de Haas oscillations.} 

\end{figure*}

\begin{figure*}[!htpb]
\renewcommand{\figurename}{Supplementary Figure}
\renewcommand{\thefigure}{\arabic{figure}.}
\centering
\includegraphics[width=\textwidth]{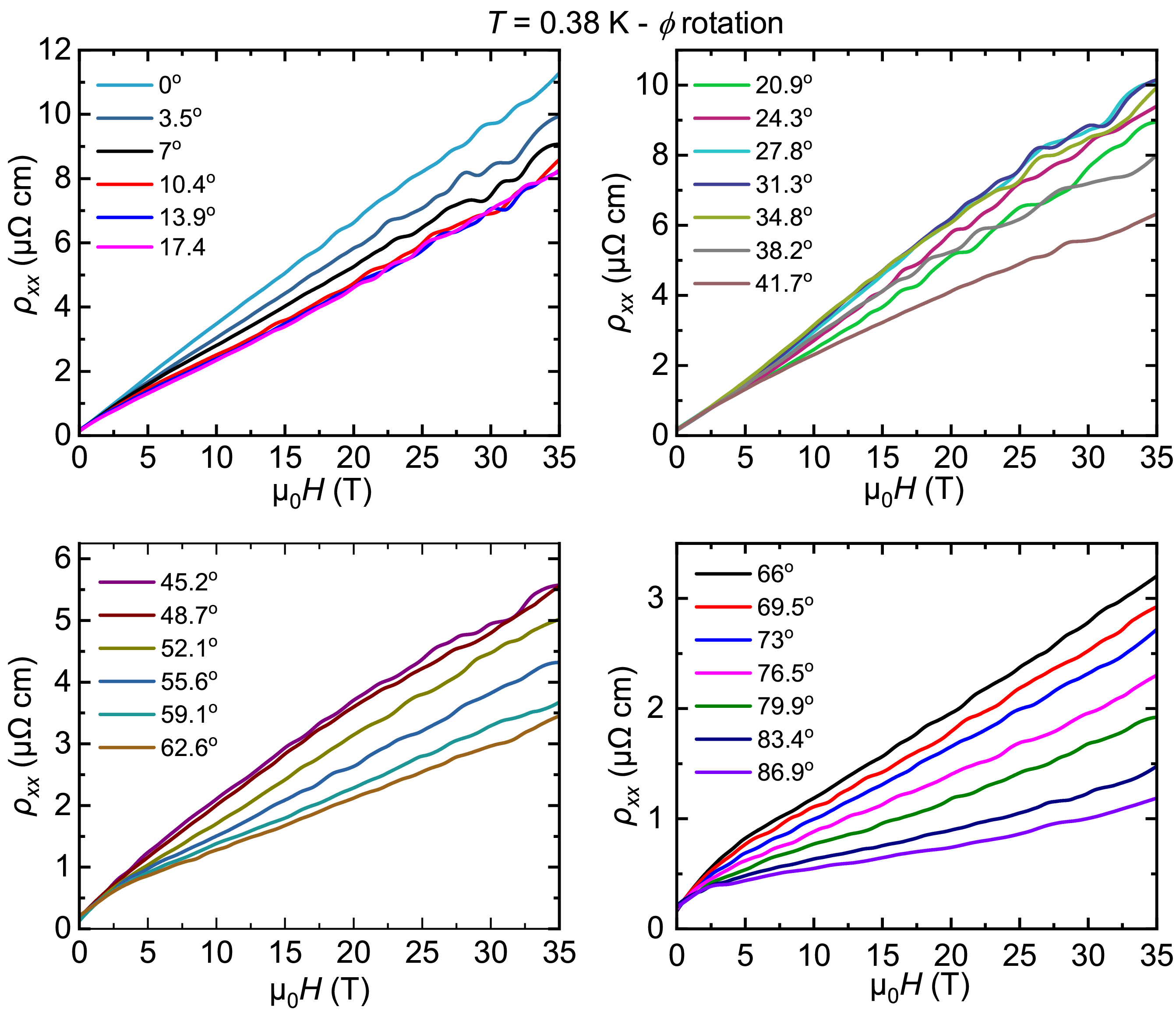}
\caption{\label{fig2:epsart}\textbf{Raw data of the linear and sublinear magnetoresistivity for several angles $\phi$.} Raw magnetoresistivity data as a function of  $\mu_0H$ for different angles $\phi$ where $\phi = 0^{\circ}$ corresponds to $I \perp \mu_0 H$, and $\phi = 90^{\circ}$ to $I \parallel \mu_0H$, at a temperature $T$ = 0.38 K.}
\end{figure*}

\begin{figure*}[!htpb]
\renewcommand{\figurename}{Supplementary Figure}
\renewcommand{\thefigure}{\arabic{figure}.}
\centering
\includegraphics[width=0.6\textwidth]{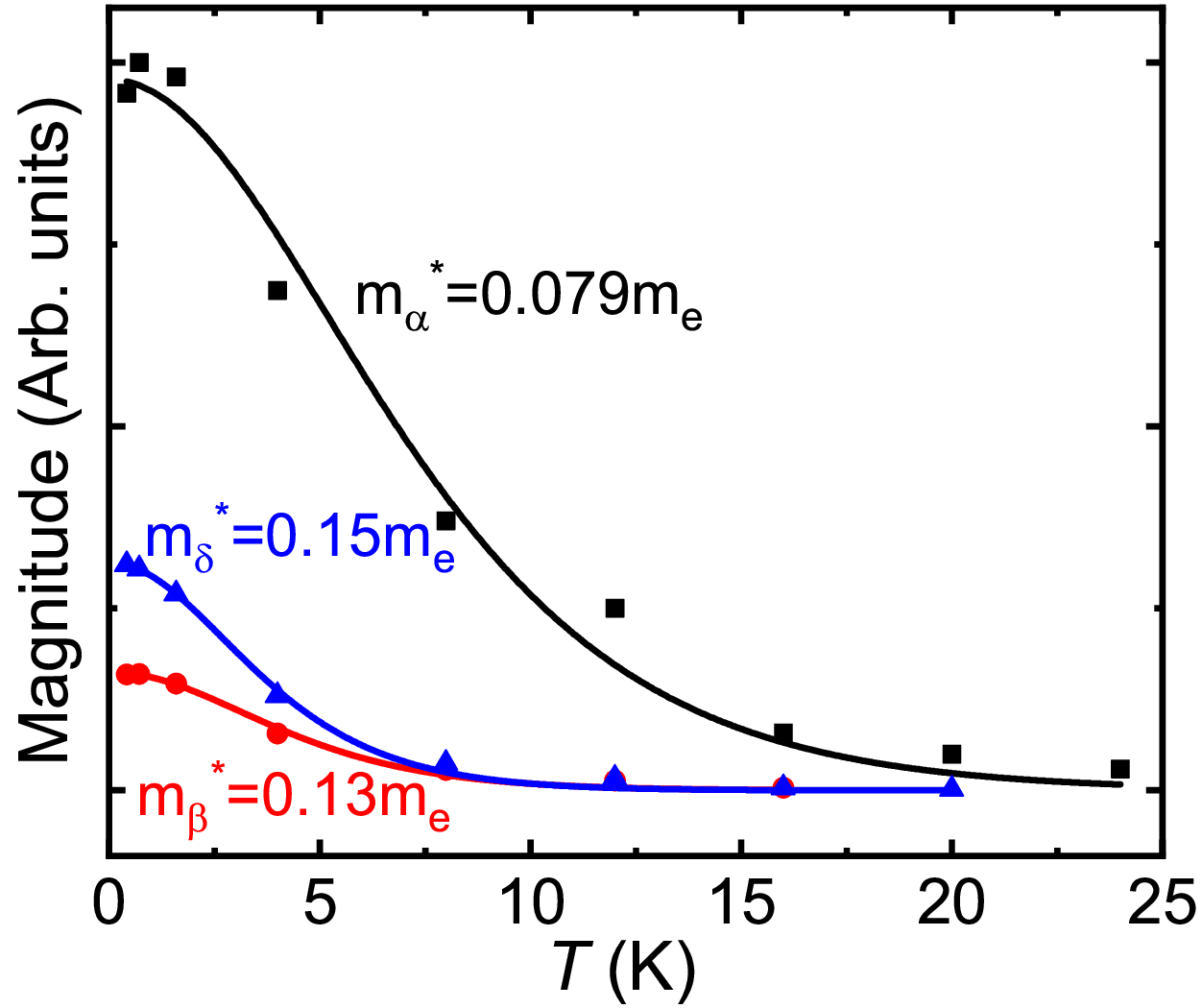}
\caption{\label{fig3:epsart}\textbf{Effective masses as extracted from the de Haas-van Alphen signal.} Magnitude of a few of the peaks observed in the Fourier transform spectra of the de Haas - van Alphen signal of Pd$_3$In$_7$ as function of the temperature $T$. The colored solid markers correspond to the experimental points, while the solid lines correspond to fittings to the Lifshitz-Kosevich temperature damping term from which we extract the effective masses m* for the different frequencies. The extracted values of m* are shown. 
}
\end{figure*}

\begin{figure*}[!htpb]
\renewcommand{\figurename}{Supplementary Figure}
\renewcommand{\thefigure}{\arabic{figure}.}
\centering
\includegraphics[width=0.8\textwidth]{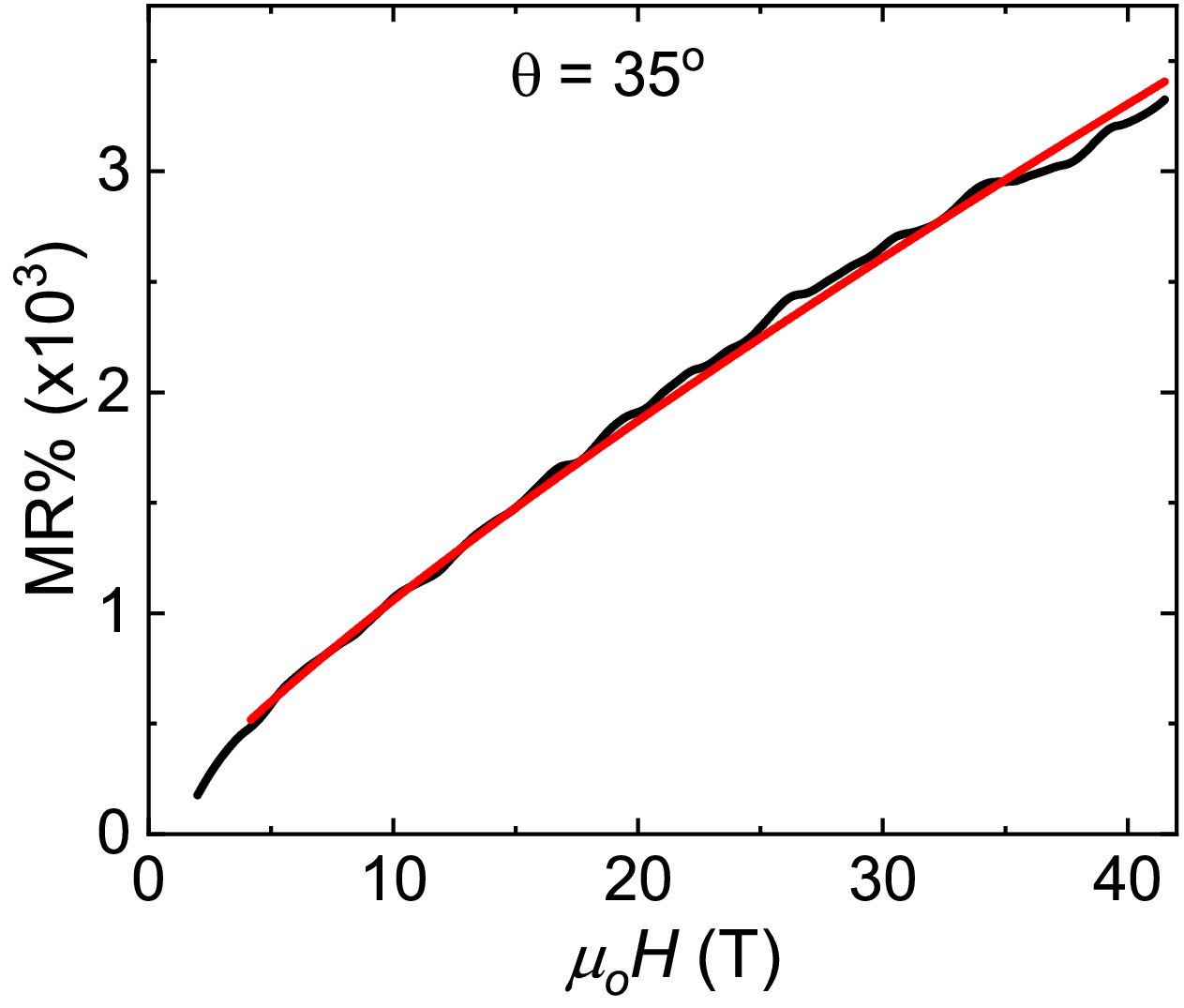}
\caption{\label{fig4:epsart}\textbf{Sublinear magnetoresistance.} MR as a function of $\mu_0 H$ for 3 T $\leqq$ $\mu_0 H$ $\leqq$  41.5 T and for $\theta = 35^\circ$. Red line is a fit to MR $= a(\mu_0 H)^n$ for $\mu_0H > $ 4 T. }
\end{figure*}

\begin{figure*}[!htpb]
\renewcommand{\figurename}{Supplementary Figure}
\renewcommand{\thefigure}{\arabic{figure}.}
\centering
\includegraphics[width=0.8\textwidth]{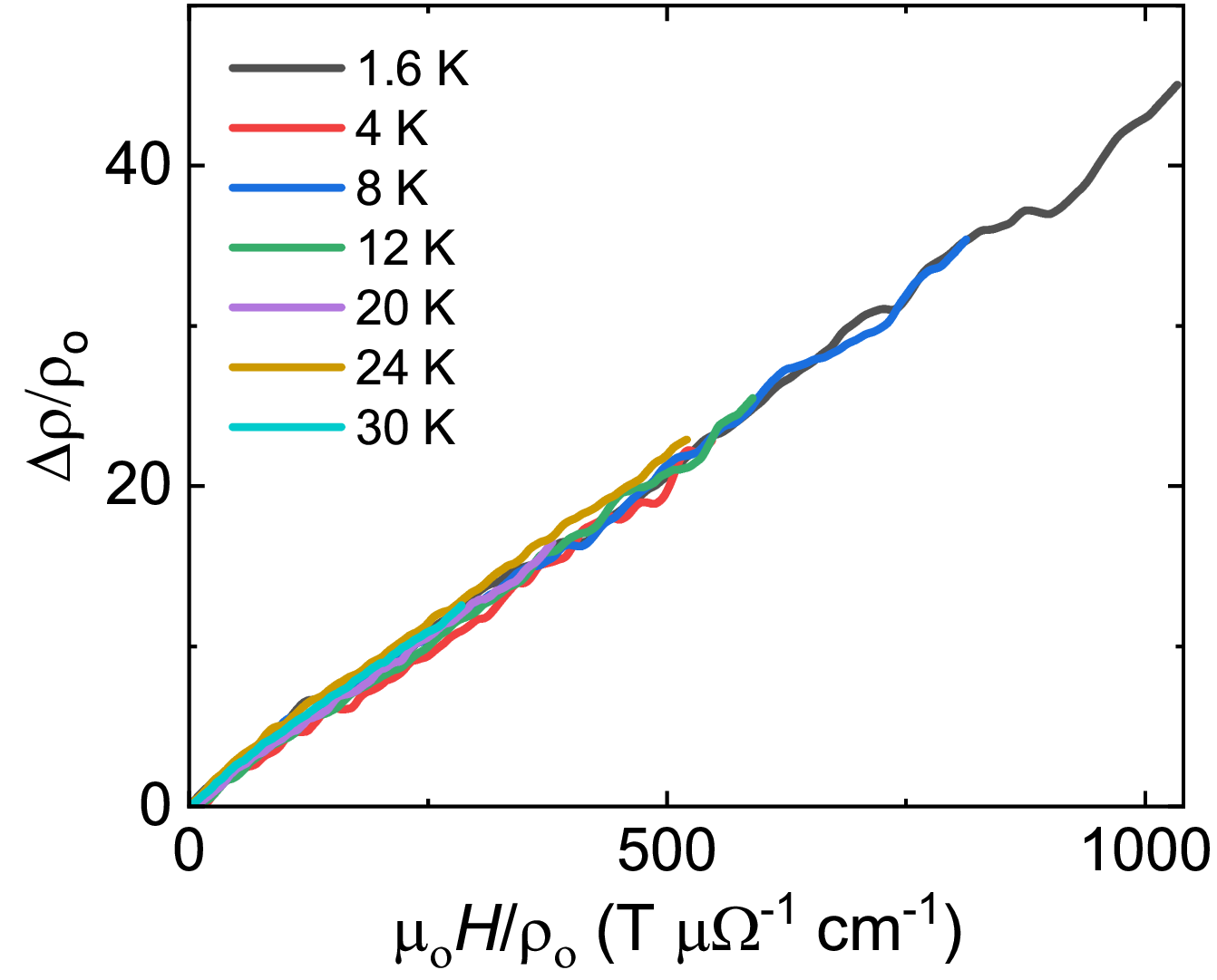}
\caption{\label{fig5:epsart}\textbf{Kohler plot at $\theta=0^\circ$.} Kohler plot of MR$(\mu_0H, T) $ indicating that the Kohler's rule is obeyed for this particular orientation. }
\end{figure*}

\begin{figure*}[!htpb]
\renewcommand{\figurename}{Supplementary Figure}
\renewcommand{\thefigure}{\arabic{figure}.}
\centering
\includegraphics[width=0.6\textwidth]{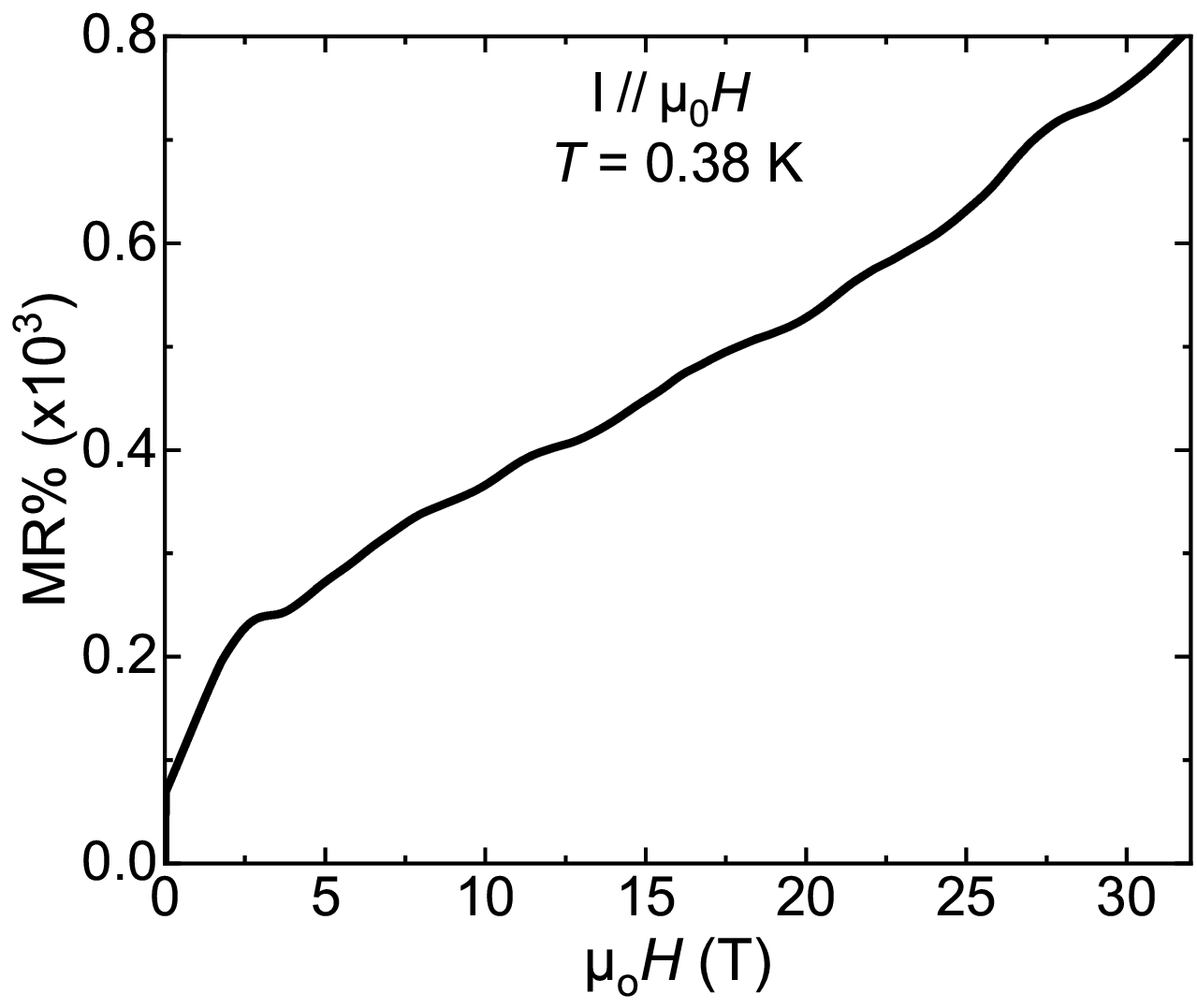}
\caption{\label{fig6:epsart}\textbf{Magnetoresistance at $\phi=90^\circ$.} Magnetoresistance as a function of the magnetic field $\mu_0H$ for a Pd$_3$In$_7$ single-crystal, when the magnetic field is alligned along the current. This trace was collected at a temperature $T = 380$ mK.}
\end{figure*}

\begin{figure*}[!htpb]
\renewcommand{\figurename}{Supplementary Figure}
\renewcommand{\thefigure}{\arabic{figure}.}
\centering
\includegraphics[width=0.8\textwidth]{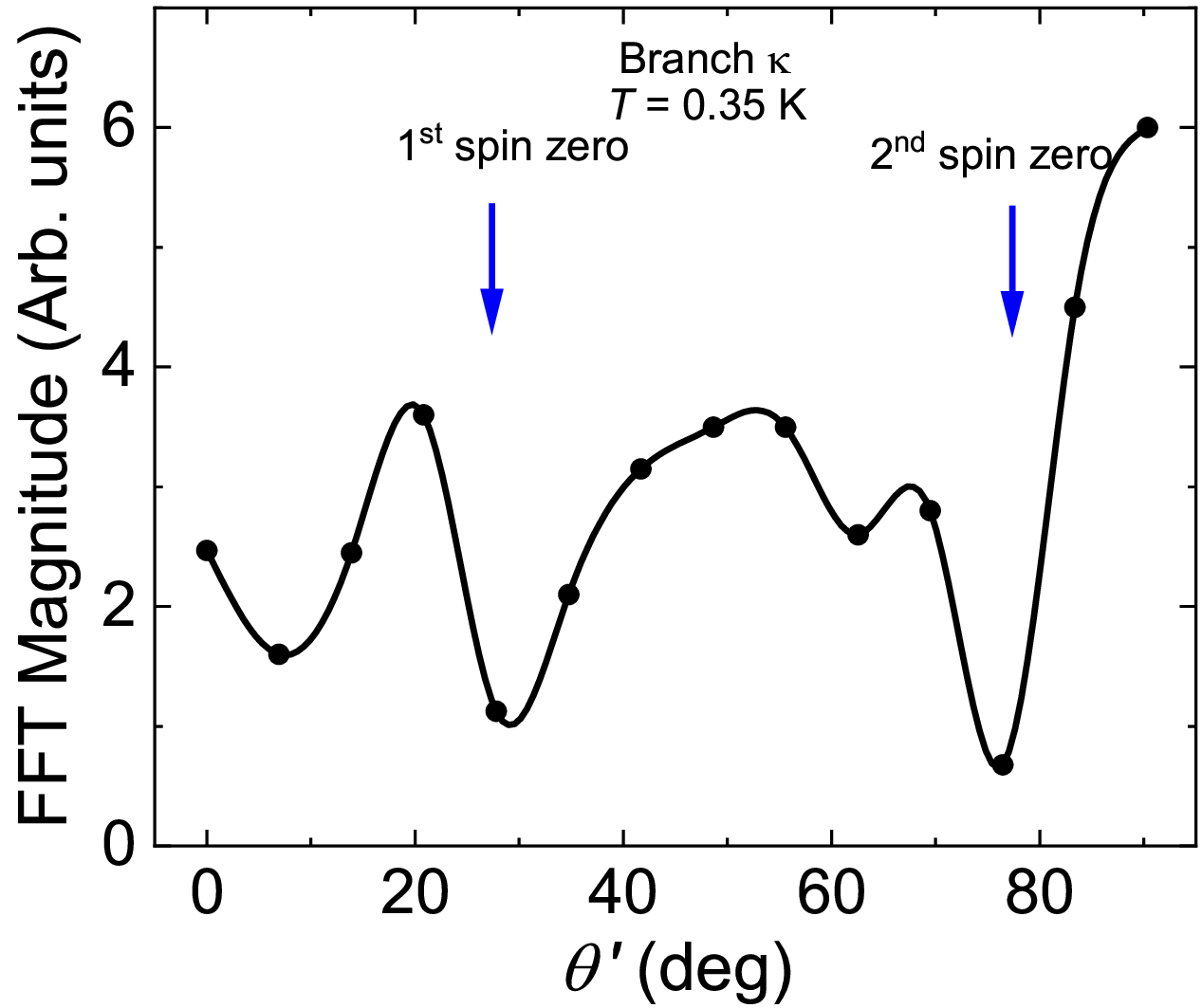}
\caption{\label{fig7:epsart}\textbf{Spin-zeros for the $\kappa$ branch.} FFT magnitude of the $\kappa$-branch frequency as a function of the angle $\theta^{\prime}$ where two spin-zeros are observed (indicated by blue arrows), leading to Land\'e $g$-factors of 9.1 and 17.64, respectively.}
\end{figure*}

\begin{figure*}[!htpb]
\renewcommand{\figurename}{Supplementary Figure}
\renewcommand{\thefigure}{\arabic{figure}.}
\centering
\includegraphics[width=0.8\textwidth]{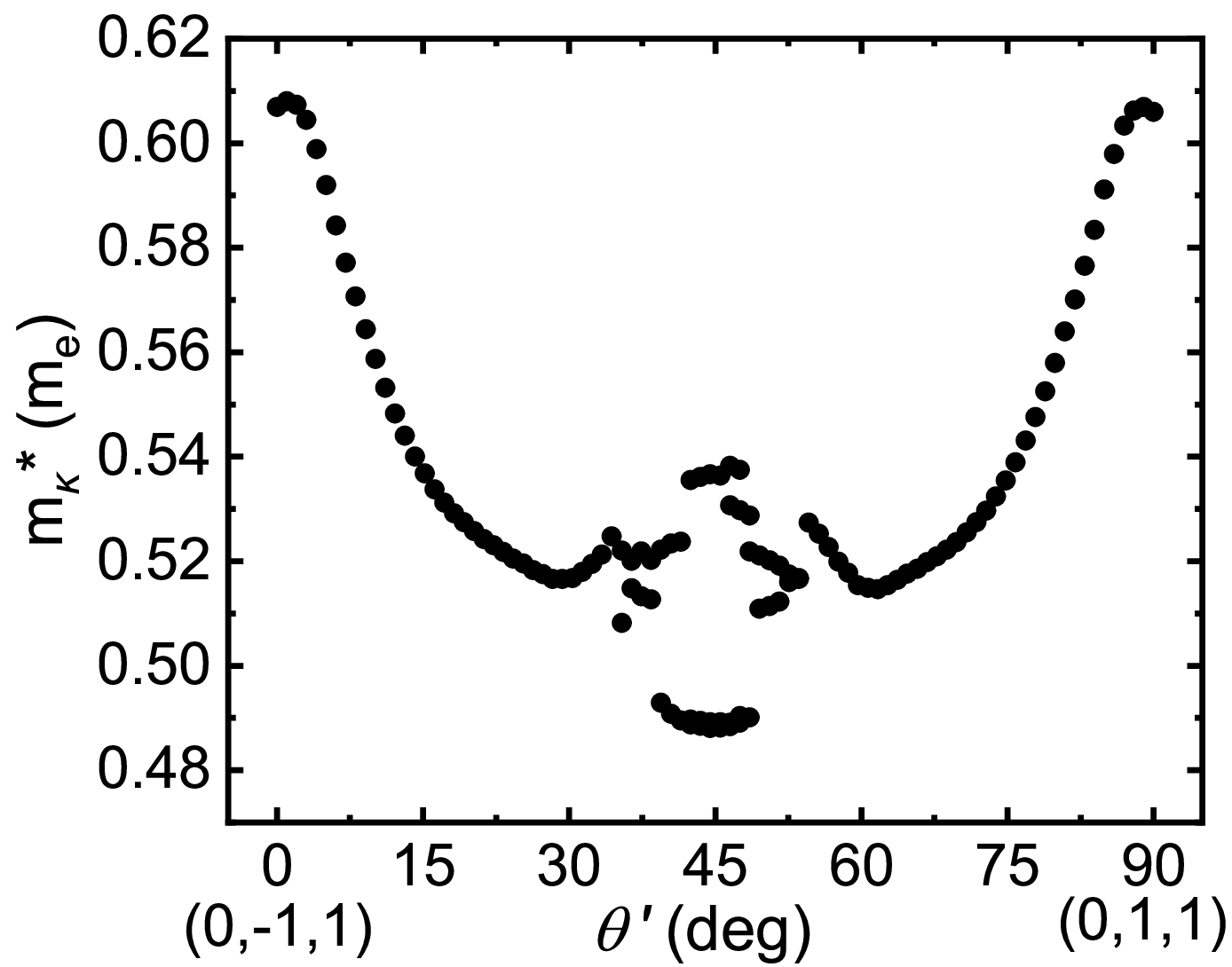}
\caption{\label{fig9:epsart}\textbf{Angular dependence of the effective mass.} Calculated effective mass $m^{\star}$ for the $\kappa$ branch as a function of $\theta^{\prime}$.}
\end{figure*}

\begin{figure*}[!htpb]
\renewcommand{\figurename}{Supplementary Figure}
\renewcommand{\thefigure}{\arabic{figure}.}
\centering
\includegraphics[width=0.8\textwidth]{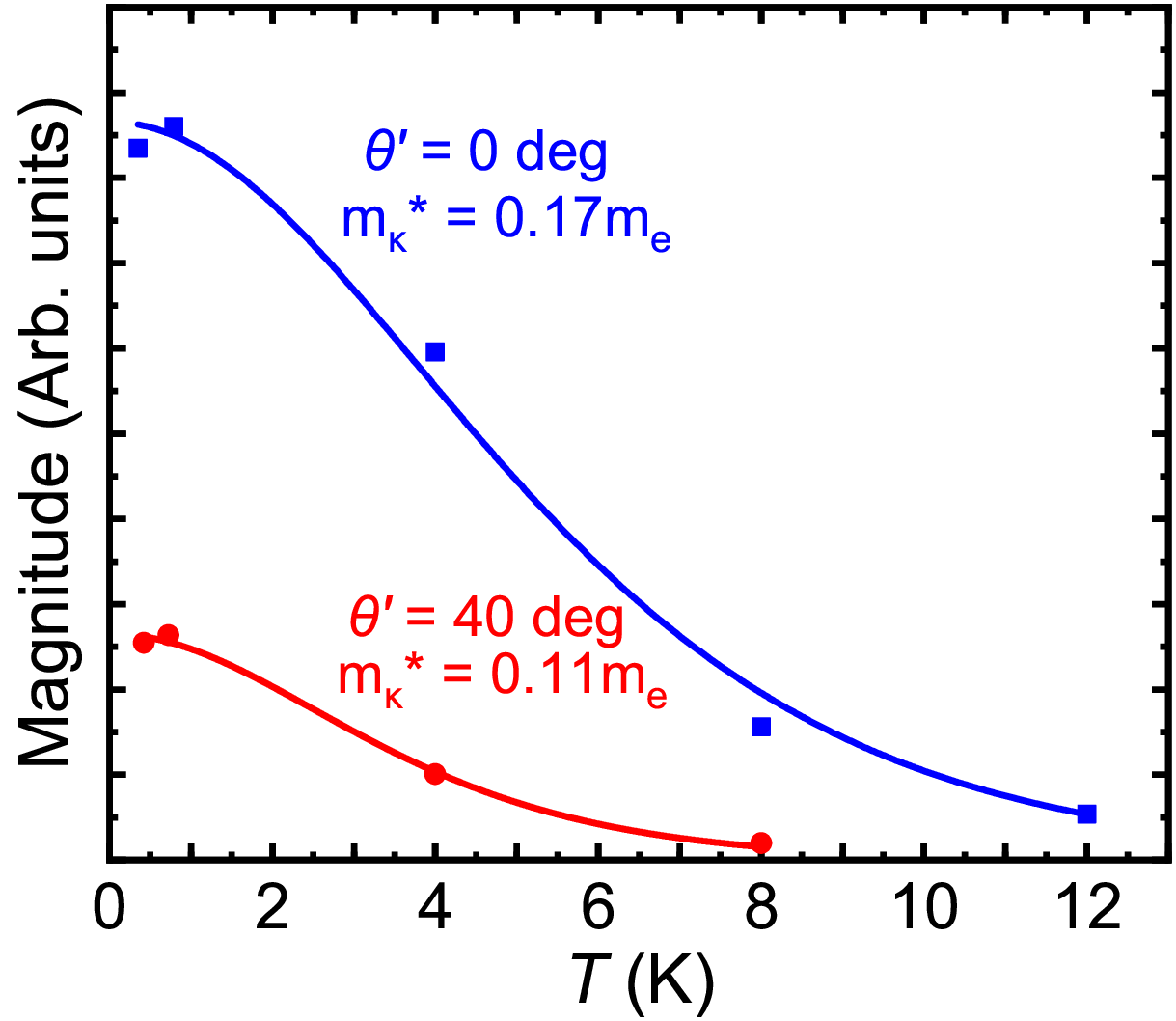}
\caption{\label{fig8:epsart}\textbf{Experimental effective masses for two orientations.} Effective mass $m_{\kappa}^{\star}$ for the $\kappa$ branch and for two different magnetic field orientations.}
\end{figure*}

\begin{figure*}[!htpb]
\renewcommand{\figurename}{Supplementary Figure}
\renewcommand{\thefigure}{\arabic{figure}.}
\centering
\includegraphics[width=0.8\textwidth]{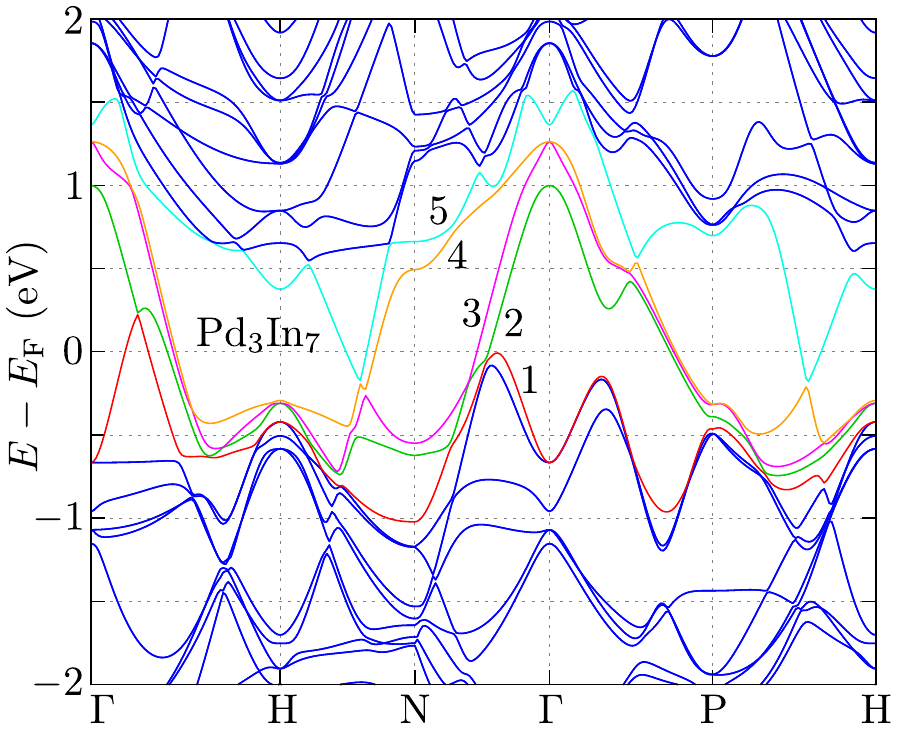}
\caption{\label{fig10:epsart}\textbf{Electronic structure of Pd$_3$In$_7$.} Electronic band structure of Pd$_3$In$_7$ including spin orbit coupling, where bands 1 - 5 yield  the different Fermi surfaces shown in Fig. 5c within the main text. Bands 1 through 3 yield hole pockets with bands 4 and 5 yielding electron-like FS sheets. Here, the bands intersecting the Fermi level are colored to associate them with the corresponding colors of the Fermi surfaces shown in the main text. 
}
\end{figure*}

\begin{figure*}[!t]
\renewcommand{\figurename}{Supplementary Figure}
\renewcommand{\thefigure}{\arabic{figure}.}
\centering
\includegraphics[width=\linewidth]{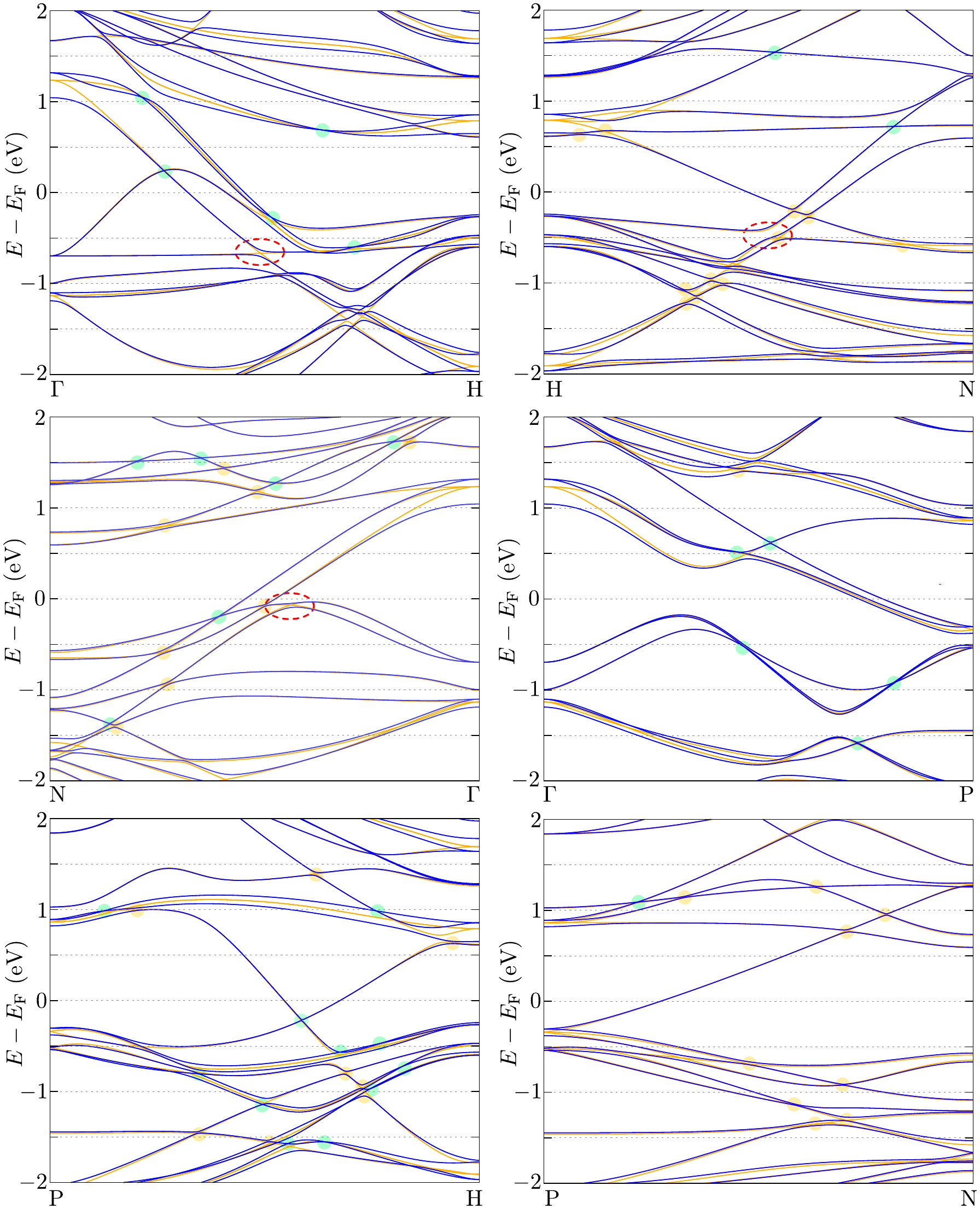}
\caption{\label{fig13:epsart}\textbf{Detailed electronic structure between high symmetry points in $k$-space.} Electronic band structure of Pd$_{3}$In$_{7}$ along all possible  directions between high symmetry points in $k$-space. Blue and orange lines correspond to calculated electronic bands that include and exclude spin-orbit coupling, respectively. Green and orange markers indicate crossings and avoided crossings, respectively. Red dashed ellipses point to the existence of sizeable gaps.
\label{fig.pdb}}
\end{figure*}


\begin{figure*}[!htpb]
\renewcommand{\figurename}{Supplementary Figure}
\renewcommand{\thefigure}{\arabic{figure}.}
\centering
\includegraphics[width=0.95\textwidth]{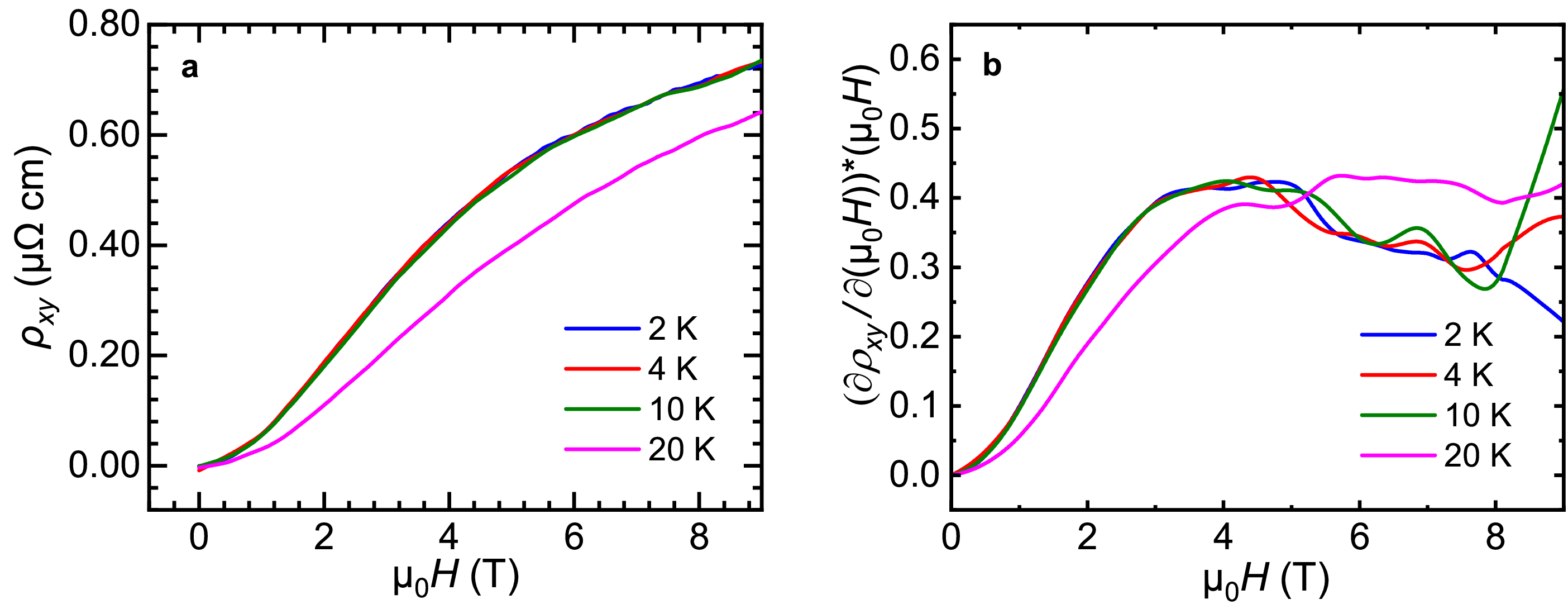}
\caption{\label{fig15:epsart}\textbf{Hall-effect for Pd$_3$In$_7$ and its derivative with respect to field.} \textbf{a}, Hall effect as a function of $\mu_0H$ at $\theta = 0^{\circ}$ for a Pd$_3$In$_7$ single-crystal and for several temperatures. 
\textbf{b}, Derivative of the Hall effect $\partial \rho_{xy}/\partial(\mu_0H) \times \mu_0H$. Notice that it is non-linear as a function of the magnetic field. The superimposed oscillatory component is the Shubnikov-de Haas effect. }
\end{figure*}

\clearpage
\textcolor{black}{
\textbf{Supplementary discussion: linear longitudinal magnetoresistivity from axial anomaly}}

We consider a simple model of a two-node Weyl semimetal given by 
\begin{eqnarray}
    H = \sum_\chi \sum_\mathbf{k} {\chi\hbar v_F (\mathbf{k}\cdot{\sigma} +t^\chi_z k_z)}
    \label{Eq:HWeyl2}
\end{eqnarray}
In the above model, $v_F$ is the Fermi velocity, $t^\chi_z$ is the tilting in the $z-$direction, ${\sigma}$ is the vector of Pauli matrices, $\chi$ indicates the chirality of the nodes ($\chi=+1$ or $\chi=-1$). We chose oppositely tilted Weyl nodes with $t^{+1}_z = -t^{-1}_z$. 

We  wish to study transport in Weyl semimetals in the limit of weak electric and magnetic fields using the quasiclassical Boltzmann approximation. Since quasiclassical Boltzmann theory is valid away from the nodal point such that $\mu^2\gg \hbar v_F^2 e \mu_0 H$, therefore, without any loss of generality we will assume that the chemical potential lies in the conduction band. The Boltzmann equation for the distribution function $f^\chi_\mathbf{k}$ is
\begin{eqnarray}
\left(\frac{\partial}{\partial t} + \dot{\mathbf{r}}^\chi\cdot \nabla_\mathbf{r}+\dot{\mathbf{k}}^\chi\cdot \nabla_\mathbf{k}\right)f^\chi_\mathbf{k} = \mathcal{I}_{\mathrm{coll}}[f^\chi_\mathbf{k}],
\label{Eq_boltz1}
\end{eqnarray}
where the collision term on the right-hand side of the equation incorporates the effects of scattering due to impurities.
In the presence of electric and magnetic fields, the dynamics of the Bloch electrons is modified as follows:
\begin{eqnarray}
\dot{\mathbf{r}}^\chi &= \mathcal{D}^\chi \left( \frac{e}{\hbar}(\mathbf{E}\times \boldsymbol{\Omega}^\chi + \frac{e}{\hbar}(\mathbf{v}^\chi\cdot \boldsymbol{\Omega}^\chi) \mu_0\mathbf{H} + \mathbf{v}_\mathbf{k}^\chi)\right) \nonumber\\
\dot{\mathbf{p}}^\chi &= -e \mathcal{D}^\chi \left( \mathbf{E} + \mathbf{v}_\mathbf{k}^\chi \times \mu_0\mathbf{H} + \frac{e}{\hbar} (\mathbf{E}\cdot\mu_0\mathbf{H}) \boldsymbol{\Omega}^\chi \right),
\end{eqnarray}
where $\mathbf{v}_\mathbf{k}^\chi$ is the semiclassical velocity, $\boldsymbol{\Omega}^\chi = -\chi \mathbf{k} /2k^3$ is the Berry curvature, and $\mathcal{D}^\chi = (1+e\mu_0\mathbf{H}\cdot\boldsymbol{\Omega}^\chi/\hbar)^{-1}$, $\mathbf{m}^\chi_\mathbf{k}$ is the orbital magnetic moment (OMM). In the presence of an external magnetic field, the OMM modifies the energy dispersion as $\epsilon^{\chi}_{\mathbf{k}}\rightarrow \epsilon^{\chi}_{\mathbf{k}} - \mathbf{m}^\chi_\mathbf{k}\cdot \mu_0\mathbf{H}$. 

The collision integral must take into account scattering internode as well as intranode scattering, and therefore, $\mathcal{I}_{\mathrm{coll}}[f^\chi_\mathbf{k}]$ can be expressed as 
\begin{eqnarray}
\mathcal{I}_{\mathrm{coll}}[f^\chi_\mathbf{k}] = \sum_{\chi'}\sum_{\mathbf{k}'} W^{\chi\chi'}_{\mathbf{k},\mathbf{k}'} (f^{\chi'}_{\mathbf{k}'} - f^\chi_\mathbf{k}),
\end{eqnarray}
where the scattering rate $W^{\chi\chi'}_{\mathbf{k},\mathbf{k}'}$ is determined by Fermi's golden rule. 

The distribution function is assumed to take the form $f^\chi_\mathbf{k} = f_0^\chi + g^\chi_\mathbf{k}$, where $f_0^\chi$ is the equilibrium Fermi-Dirac distribution function and $g^\chi_\mathbf{k}$ indicates the deviation from equilibrium. 
In the steady state, the Boltzmann equation (Eq.~\ref{Eq_boltz1}) becomes
\begin{eqnarray}
\left[\left(\frac{\partial f_0^\chi}{\partial \epsilon^\chi_\mathbf{k}}\right) \mathbf{E}\cdot \left(\mathbf{v}^\chi_\mathbf{k} + \frac{e\mathbf{B}}{\hbar} (\boldsymbol{\Omega}^\chi\cdot \mathbf{v}^\chi_\mathbf{k}) \right)\right]
= -\frac{1}{e \mathcal{D}^\chi}\sum\limits_{\chi'}\sum\limits_{\mathbf{k}'} W^{\chi\chi'}_{\mathbf{k}\mathbf{k}'} (g^\chi_{\mathbf{k}'} - g^\chi_\mathbf{k})
 \label{Eq_boltz2}
\end{eqnarray}
The deviation $g^\chi_\mathbf{k}$ is assumed to be linearly proportional to $\mathbf{E}$
\begin{eqnarray}
g^\chi_\mathbf{k} = e \left(-\frac{\partial f_0^\chi}{\partial \epsilon^\chi_\mathbf{k}}\right) \mathbf{E}\cdot \boldsymbol{\Lambda}^\chi_\mathbf{k}
\end{eqnarray}
We fix the direction of the applied external electric field to be along $+\hat{z}$, i.e., $\mathbf{E} = E\hat{z}$. Therefore, only ${\Lambda}^{\chi z}_\mathbf{k}\equiv {\Lambda}^{\chi}_\mathbf{k}$, is relevant to us. Further, the magnetic field is parallel to the $z-$axis as well.

Keeping terms only up to linear order in the electric field, Eq.~\ref{Eq_boltz2} takes the following form 
\begin{eqnarray}
\mathcal{D}^\chi \left[v^{\chi z}_{\mathbf{k}} + \frac{e B}{\hbar}  (\boldsymbol{\Omega}^\chi\cdot \mathbf{v}^\chi_\mathbf{k})\right] = \sum\limits_{\eta}\sum\limits_{\mathbf{k}'} W^{\eta\chi}_{\mathbf{k}\mathbf{k}'} (\Lambda^{\eta}_{\mathbf{k}'} - \Lambda^\chi_\mathbf{k})
\label{Eq_boltz3}
\end{eqnarray} 
We now define the valley scattering rate as follows:
\begin{eqnarray}
\frac{1}{\tau^\chi_\mathbf{k}} = \mathcal{V} \sum\limits_{\eta} \int{\frac{d^3 \mathbf{k}'}{(2\pi)^3} (\mathcal{D}^\eta_{\mathbf{k}'})^{-1} W^{\eta\chi}_{\mathbf{k}\mathbf{k}'}}
\label{Eq_tau11}
\end{eqnarray}
The radial integration is simplified due to the delta-function in the collision integral. 

Substituting the scattering rate in the above equation, we have 
\begin{eqnarray}
\frac{1}{\tau^\chi_\mathbf{k}} = \frac{\mathcal{V}N}{8\pi^2 \hbar} \sum\limits_{\eta} (U^{\chi\eta})^2 \iiint{q^2 \sin \theta' \mathcal{G}^{\chi\eta}(\theta,\phi,\theta',\phi') \delta(\epsilon^{\eta}_{\mathbf{q}}-\epsilon_F)(\mathcal{D}^\eta_{\mathbf{q}})^{-1}dq d\theta'd\phi'},\nonumber\\
\label{Eq_tau1}
\end{eqnarray}
where $N$ now indicates the total number of impurities, and $ \mathcal{G}^{\chi\eta}(\theta,\phi,\theta',\phi') = (1+\chi\eta(\cos\theta \cos\theta' + \sin\theta\sin\theta' \cos(\phi-\phi')))$ is the Weyl chirality factor defined by the overlap of the wavefunctions. The Fermi wavevector contour $k^\chi$ is evaluated by equating the energy expression with the Fermi energy.
The three-dimensional integral in Eq.~\ref{Eq_tau1} is  reduced to just integration in $\phi'$ and $\theta'$. The scattering time ${\tau^\chi_\mathbf{k}}$ depends on the chemical potential ($\mu$), and is a function of the angular variables $\theta$ and $\phi$. 

\begin{eqnarray}
\frac{1}{\tau^\chi_\mu(\theta,\phi)} = \mathcal{V} \sum\limits_{\eta} \iint{\frac{\beta^{\chi\eta}(k')^3}{\mathrm{abs}(\mathbf{v}^\eta_{\mathbf{k}'}\cdot \mathbf{k}'^\eta)}\sin\theta'\mathcal{G}^{\chi\eta}(\mathcal{D}^\eta_{\mathbf{k}'})^{-1} d\theta'd\phi'},
\label{Eq_tau2}
\end{eqnarray}
\begin{figure}
    \centering
    \renewcommand{\figurename}{Supplementary Figure}
    \renewcommand{\thefigure}{\arabic{figure}.}
    \includegraphics[width =\columnwidth]{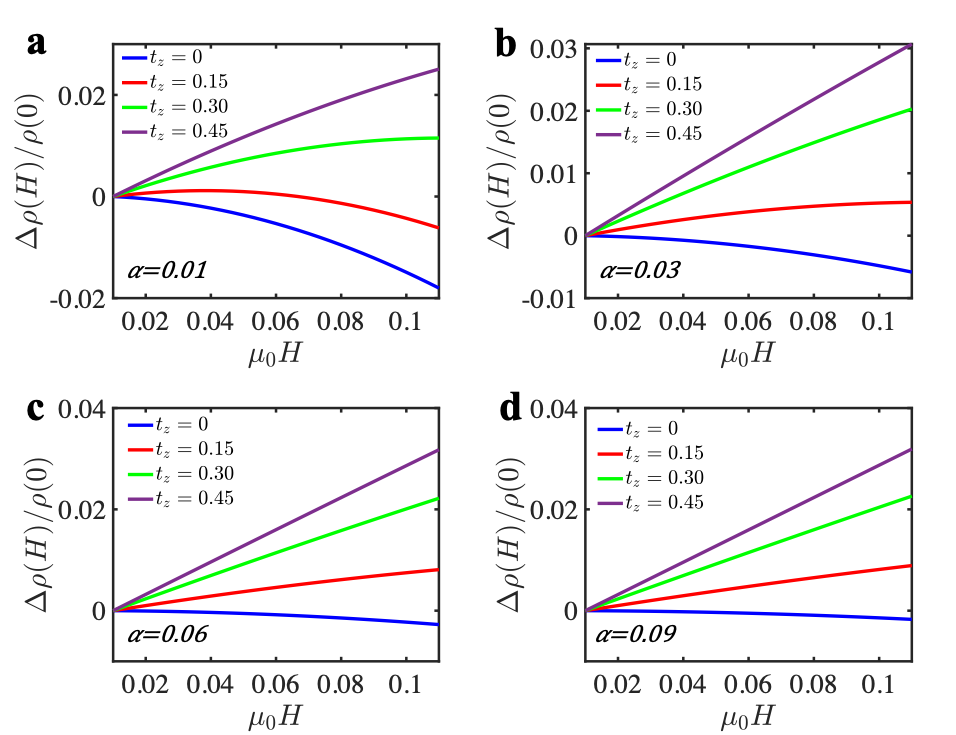}
    \caption{\textbf{Modeled longitudinal magnetoresistance.} Longitudinal magnetoresistance for the model presented in Eq.~\ref{Eq:HWeyl2}. The dimensionless intervalley scattering strength is represented by $\alpha$, and $t_z$ is the Dirac cone tilt parameter. A finite intervalley scattering reverses the sign of the magnetoresistance while a finite tilting of the cone results in linear magnetoresistivity.  }
    \label{fig:magres_model}
\end{figure}

The Boltzmann equation (Eq~\ref{Eq_boltz3}) now becomes 
\begin{eqnarray}
h^\chi_\mu(\theta,\phi) + \frac{\Lambda^\chi_\mu(\theta,\phi)}{\tau^\chi_\mu(\theta,\phi)} =\mathcal{V}\sum_\eta \iint {\frac{\beta^{\chi\eta}(k')^3}{\mathrm{abs}(\mathbf{v}^\eta_{\mathbf{k}'}\cdot \mathbf{k}'^\eta)} \sin\theta'\mathcal{G}^{\chi\eta}(\mathcal{D}^\eta_{\mathbf{k}'})^{-1}\Lambda^\eta_{\mu}(\theta',\phi') d\theta'd\phi'}\nonumber\\
\label{Eq_boltz4}
\end{eqnarray}
We assume the following ansatz for $\Lambda^\chi_\mu(\theta,\phi)$
\begin{eqnarray}
\Lambda^\chi_\mu(\theta,\phi) = (\lambda^\chi - h^\chi_\mu(\theta,\phi) + a^\chi \cos\theta +b^\chi \sin\theta\cos\phi + c^\chi \sin\theta\sin\phi)\tau^\chi_\mu(\theta,\phi),
\label{Eq_Lambda_1}
\end{eqnarray}
where we solve for the eight unknowns ($\lambda^{\pm 1}, a^{\pm 1}, b^{\pm 1}, c^{\pm 1}$). The L.H.S in Eq.~\ref{Eq_boltz4} simplifies to $\lambda^\chi + a^\chi \cos\theta + b^\chi \sin\theta\cos\phi + c^\chi \sin\theta\sin\phi$. The R.H.S of Eq.~\ref{Eq_boltz4} simplifies to
\begin{eqnarray}
\mathcal{V}\sum_\eta \beta^{\chi\eta} \iint &f^{\eta} (\theta',\phi') \mathcal{G}^{\chi\eta} (\lambda^\eta - h^\eta_\mu(\theta',\phi') + a^\eta \cos\theta' +\nonumber\\
	&b^\eta \sin\theta'\cos\phi' + c^\eta \sin\theta'\sin\phi')d\theta'd\phi',
	\label{Eq_boltz5_rhs}
\end{eqnarray}
where the function
\begin{eqnarray}
f^{\eta} (\theta',\phi') = \frac{(k')^3}{\mathrm{abs}(\mathbf{v}^\eta_{\mathbf{k}'}\cdot \mathbf{k}'^\eta)} \sin\theta' (\mathcal{D}^\eta_{\mathbf{k}'})^{-1} \tau^\chi_\mu(\theta',\phi')
\label{Eq_f_eta}
\end{eqnarray}
The above equations, when written down explicitly take the form of seven simultaneous equations to be solved for eight variables. The final constraint comes from charge conservation 
\begin{eqnarray}
\sum\limits_{\chi}\sum\limits_{\mathbf{k}} g^\chi_\mathbf{k} = 0
\label{Eq_sumgk}
\end{eqnarray}
Eq.~\ref{Eq_Lambda_1}, Eq.~\ref{Eq_boltz5_rhs}, Eq.~\ref{Eq_f_eta} and Eq.~\ref{Eq_sumgk} are solved together with Eq~\ref{Eq_tau2}, simultaneously for the eight unknowns ($\lambda^{\pm 1}, a^{\pm 1}, b^{\pm 1}, c^{\pm 1}$). All the two dimensional integrals with respect to \{$\theta'$, $\phi'$\}, and the solution of the simultaneous equations are performed numerically. 

In Fig.~\ref{fig:magres_model} we plot the longitudinal magnetoresistance for the model presented in Eq.~\ref{Eq:HWeyl2}. We find that a finite intervalley scattering reverses the sign of the magnetoresistance while a finite tilting of the cone results in linear magnetoresistance. Although results are presented for a type-I tilted Weyl semimetal, we expect the qualitative features to remain unchanged even for higher tilt values, i.e, type-II Weyl semimetal.